\documentclass[10pt]{article}
\usepackage[svgnames]{xcolor}
\usepackage[nofonts]{dgleich-common}
\usepackage[T1]{fontenc}
\RequirePackage[scaled=0.8]{helvet}%
\RequirePackage[scaled=0.75]{beramono}%
\def\lstbasicfont{\fontfamily{fvm}\selectfont}
\makeatletter  
\setboolean{@dgleich@sanssmallcaps}{true}
\newcommand{\resetcaptions}{%
    \let\caption\@tufte@orig@caption%
    \let\label\@tufte@orig@label%
}
\makeatother
\usepackage{dgleich-math}

\newcommand{\secref}[1]{\hyperref[#1]{section~\oldstylenums{\ref{#1}}}}
\newcommand{\tabref}[1]{\hyperref[#1]{table~\oldstylenums{\ref{#1}}}}
\newcommand{\Tabref}[1]{\hyperref[#1]{Table~\oldstylenums{\ref{#1}}}}
\newcommand{\figref}[1]{\hyperref[#1]{figure~\oldstylenums{\ref{#1}}}}
\makeatletter
\renewcommand\normalsize{%
   \@setfontsize\normalsize\@xpt{14}%
   \abovedisplayskip 10\p@ \@plus2\p@ \@minus5\p@
   \abovedisplayshortskip \z@ \@plus3\p@
   \belowdisplayshortskip 6\p@ \@plus3\p@ \@minus3\p@
   \belowdisplayskip \abovedisplayskip
   \let\@listi\@listI}
\normalsize
\renewcommand\small{%
   \@setfontsize\small\@ixpt{12}%
   \abovedisplayskip 8.5\p@ \@plus3\p@ \@minus4\p@
   \abovedisplayshortskip \z@ \@plus2\p@
   \belowdisplayshortskip 4\p@ \@plus2\p@ \@minus2\p@
   \def\@listi{\leftmargin\leftmargini
               \topsep 4\p@ \@plus2\p@ \@minus2\p@
               \parsep 2\p@ \@plus\p@ \@minus\p@
               \itemsep \parsep}%
   \belowdisplayskip \abovedisplayskip
}
\renewcommand\footnotesize{%
   \@setfontsize\footnotesize\@viiipt{10}%
   \abovedisplayskip 6\p@ \@plus2\p@ \@minus4\p@
   \abovedisplayshortskip \z@ \@plus\p@
   \belowdisplayshortskip 3\p@ \@plus\p@ \@minus2\p@
   \def\@listi{\leftmargin\leftmargini
               \topsep 3\p@ \@plus\p@ \@minus\p@
               \parsep 2\p@ \@plus\p@ \@minus\p@
               \itemsep \parsep}%
   \belowdisplayskip \abovedisplayskip
}
\renewcommand\scriptsize{\@setfontsize\scriptsize\@viipt\@viiipt}
\renewcommand\tiny{\@setfontsize\tiny\@vpt\@vipt}
\renewcommand\large{\@setfontsize\large\@xipt{15}}
\renewcommand\Large{\@setfontsize\Large\@xiipt{16}}
\renewcommand\LARGE{\@setfontsize\LARGE\@xivpt{18}}
\renewcommand\huge{\@setfontsize\huge\@xxpt{30}}
\renewcommand\Huge{\@setfontsize\Huge{24}{36}}
\makeatother

\usepackage{tabularx}
\usepackage{booktabs}
\usepackage{graphicx}

\usenatbib
\usehyperref

\title{\texttt{\fontsize{12}{16}\selectfont fauci-email}: a \texttt{\fontsize{12}{16}\selectfont json} digest of Anthony Fauci's released emails}
\author{Austin R. Benson, Nate Veldt, David F. Gleich}
\date{July 2021}

\newcommand{\vol}{\mathop{\text{vol}}}
\newcommand{\cut}{\mathop{\text{cut}}}
\newcommand{\ncut}{\mathop{\text{ncut}}}
\newcommand{\cond}{\mathop{\phi}}
\newcommand{\Sbar}{\bar{S}}
\begin{document}

\maketitle

%\section*{TODO}
%\begin{itemize}
    %\item Austin - Description of automated parsing
    %\item Nate - some simple cooked hyeprgraph example with st-cuts...
    %\item Austin - will get David a sender-by-receiver-by-date-by-wordembedding tensor for analysis...
    %\item David / Austin - some tensor analysis
    %\item David / Austin - get a 77x77x212x102 'enron-email-like tensor' (sender-receiver-word-time)
    %\item Austin - try something with i,j,k = sender, receiver, cc emails. (Filter down to largest scc in all modes...??) - hypergraph centrality.
    %\item Austin - will run SVD on word embeddings...
    %\item Austin - some text analysis with word embeddings
    %\item David - will writeup mincond vs. ncut example -- 
    %\item David - writeup simple st-cut stuff...
    %\item David - temporal graph stuff...
    %\item David/Nate - give exact modularity scores.
    %\item David - list of graphs and data constructed... 
    
%\end{itemize}

\begin{abstract}
    A collection of over 3000 pages of emails sent by Anthony Fauci and his staff were released in an effort to understand the United States government response to the COVID-19 pandemic. We describe how this email data was translated into a resource consisting of  \texttt{json} files that make many future studies easy. Findings from our processed data include (i) successful organizational partitions using the simple mincut techniques in Zachary's karate club methodology, (ii) a natural example where the normalized cut and minimum conductance set are extremely different, and (iii) organizational groups identified by optimal modularity clusters that illustrate a working hierarchy.  These example uses suggest the data will be useful for future research and pedagogical uses in terms of human and system behavioral interactions. We explain a number of ways to turn email information into a network, a hypergraph, a temporal sequence, and a tensor for subsequent analysis as well as a few examples of such analysis. 
\end{abstract}

\section{Data Summary and key findings}
\label{sec:findings}

Fauci's email release~\cite{Leopold-2021-fauci-emails} includes approximately 1289 email threads with 2761 emails including 101 duplicate emails among the threads.\footnote{These counts are exact only given the precise details of our data conversion strategy including what is retained and excluded, see more below; independent parsing and analysis may show slightly different counts.} Each email thread begins with an email from Fauci and the thread is the chain of emails underlying his reply. There are 410 length 1 threads of outgoing mail only.  Each email includes partially redacted text and is time-stamped, albeit from a mixture of time-zones that may not always be listed. 

These raw data can be analyzed as a \emph{social network or graph}, \emph{a temporal graph}, \emph{a hypergraph}, or \emph{a tensor}. The most closely related existing dataset is the Enron email dump~\cite{Cohen-2004-enron}.  For Fauci's email, we discuss a number of interesting findings in the data and provide them as an easy-to-use resource for continued exploration by others in the field. As was also the case for the Enron email dataset, there may be future releases of this data that correct errors. Our current parsing, for instance, has numerous OCR errors in the text pieces. 

The graphs, networks, and hypergraphs that result from these data are small compared with the size of many modern datasets, yet they are not so small as to permit trivial analysis. This renders them a rich setting to investigate what can be ascertained from the data. Because the original emails are available, many of these findings are easy to assess in the documents themselves to understand where various graph features arise. 

This goal of this manuscript is more akin to a data manual instead of an article that supports conclusions. We intended to highlight interesting findings of the data (\hyperref[sec:stcut-analysis]{sections~\ref{sec:stcut-analysis}--\ref{sec:modularity}}) and demonstrate a variety of uses (\secref{sec:examples}). The processed datasets we have are available on github:
\begin{center}
    \url{https://github.com/nveldt/fauci-email}
\end{center}

\begin{compactitem}
\item The main \texttt{json} digest derived from \citet{Leopold-2021-fauci-emails}, which has senders and receivers of Fauci's email threads canonically labeled in an easy-to-process format (\secref{sec:processed}).
\item Five graphs derived from the data from the data (\secref{sec:graphs} and \tabref{tab:graphs}) ranging from 46 to 869 vertices. 
\item A hypergraph derived from the emails themselves (\secref{sec:hypergraph}) with 233 nodes and 254 hyperedges. 
\item A temporal sequence of adjacency matrices over 100 days from those 77 people where information can flow among all individuals in a temporally consistent sequence (\secref{sec:temporal}).
\item A tensor projection of the data designed to highlight the role of email carbon copy (CC) networks suitable for hypergraph centrality studies (\secref{sec:tensor}) as well as a tensor representation of the data as sender, receiver, time, and word.
\end{compactitem}

\marginnote{%
\noindent\textbf{Summary of key people.}
We provide a briefly annotated list of key individuals to help contextualize some of our results. 

\begin{description}
\item[Anthony Fauci] Head of the National Institute of Allergy and Infectious Disease (NIAID), a group within the US National Institutes of Health (NIH). 
\item[Patricia Conrad] Fauci's key special assistant and frequent email proxy.
\item[Francis Collins] Head of the US National Institutes of Health (NIH), the NIH are an organizational division of the US Department of Health and Human Services (HHS). 
\item[Robert Redfield] Head of the US Centers for Disease Control (CDC), the CDC are another organizational division of HHS.  
\item[Alex Azar] Secretary (head) of the US Department of Health and Human Services (HHS), part of the president's cabinet. 
\item[Robert Kadlec] Assistant Secretary of Preparedness and Response for HHS.
\item[Deborah Birx] The Coronavirus Response Coordinator appointed by the US President and a member of the Coronavirus Task Force. 
\item[Jennifer Routh] Science communication editor in the NIAID division of the NIH. 
\item[Greg Folkers] Anthony Fauci's chief of staff. 
\end{description}
}

%\normalsize 

\subsection{Minimum Cut analysis}
\label{sec:stcut-analysis}
An early and well-known example of social network analysis was the study of a karate club by \citet{Zachary-1977-flow}. A simple minimum cut analysis of this network predicted a future division of the club into two groups.  We also found minimum cut analyses effective for weighted networks derived from the email exchanges. For instance, consider an undirected, weighted network based on senders and receivers of any email with less than 5 recipients, where edges and edge weights indicate the maximum number of emails sent along that edge or received along that edge (this is the \texttt{tofrom-nofauci-nocc} network in our detailed description, \secref{sec:graphs}). We also remove Fauci from this network, which is done both because Fauci is connected to almost everyone due to how the data were collected and also because theories of structural holes in social networks suggest more meaningful analysis with Fauci removed~\cite{Burt-1995-structural}. Finally, we examine the minimum cut between Francis Collins (head of the NIH) and Patricia Conrad (Fauci's assistant). This cut roughly bisects the network into two pieces as shown in \figref{fig:stcut-conrad-collins}. There are 16 edges cut listed in the figure. This cut is largely preserved under multiple perturbations of the network structure (e.g., considering hypergraph projections, including additional emails with more recipients weighted to scale edge importance with recipient list size). 

An interesting node in the cut list is Sheila Kaplan, who is a New York Times Reporter. Her interactions with Collins and Conrad revolved around the New York Times desire to interview Fauci for an article around March 16-18 -- this involves Kaplan discussing the issue with the NIH Office of Communication.

\begin{fullwidthfigure}
    \resetcaptions
    %\centering
    \(\vcenter{\hbox{\text{\includegraphics[width=0.35\linewidth]{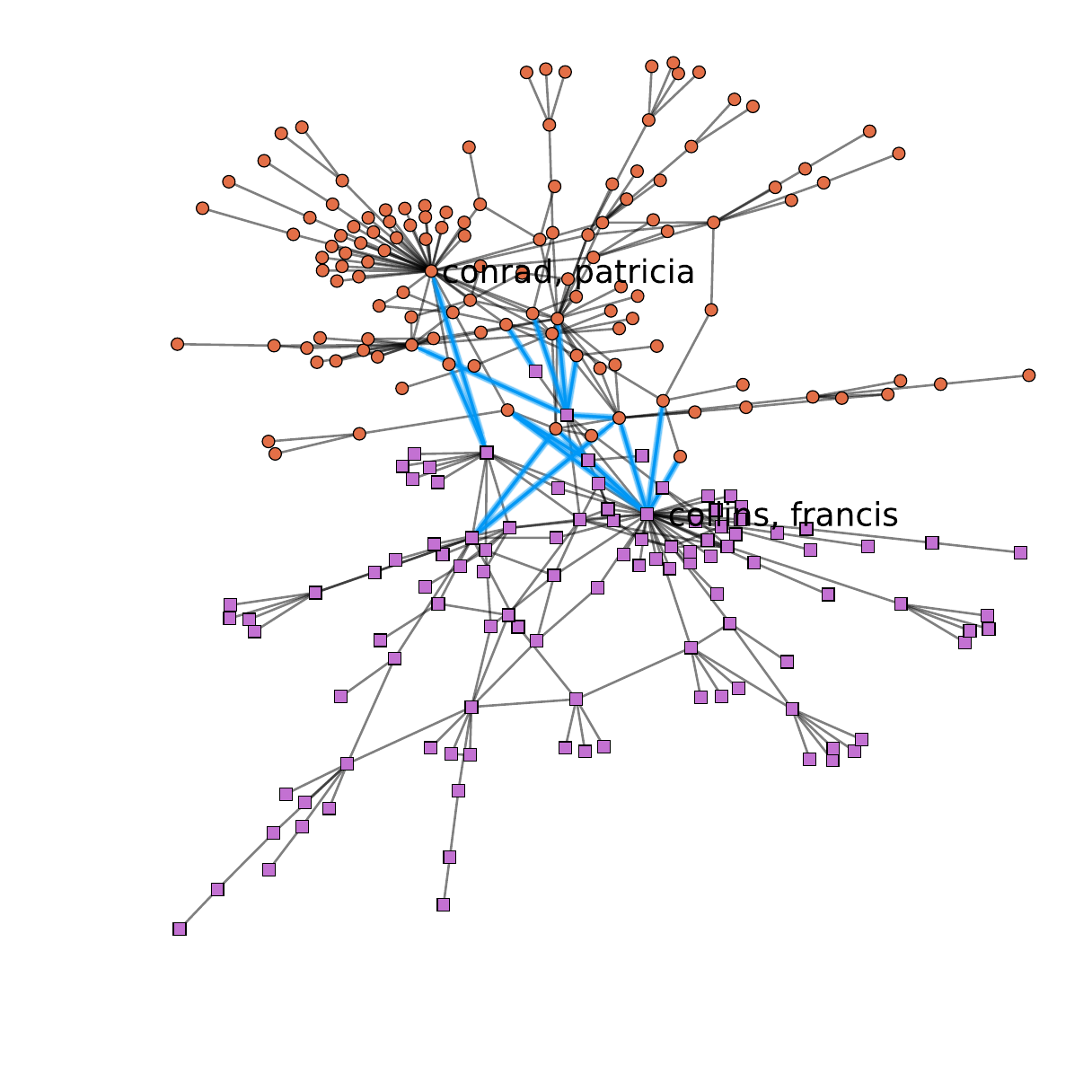}}}}\)%
    \(\vcenter{\hbox{\text{\includegraphics[width=0.35\linewidth]{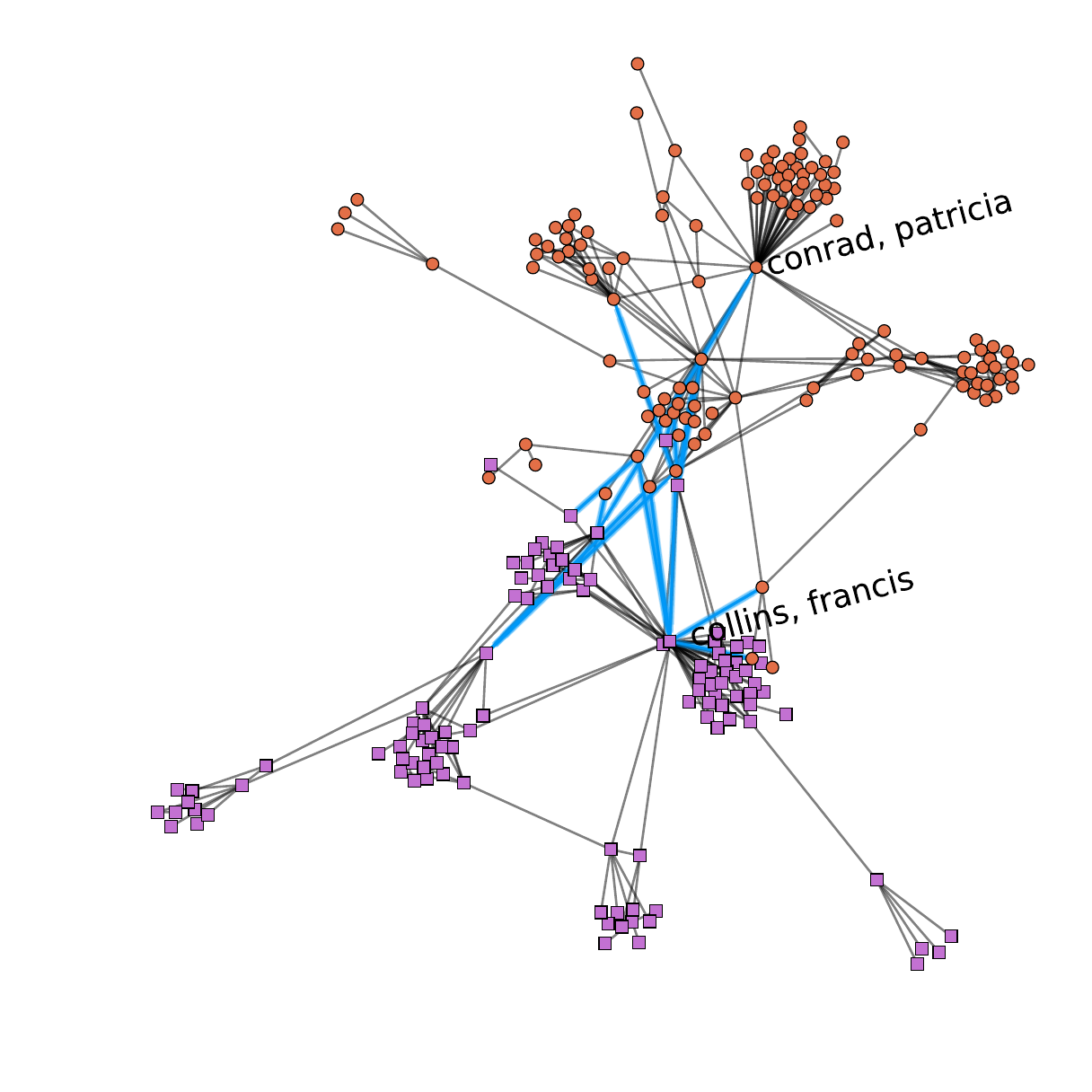}}}}\)%
    \begin{minipage}[m]{0.3\linewidth}%
    \centering%
    \emph{Cut edges}\\
    \medskip
    
    \footnotesize%
    \begin{tabularx}{\linewidth}{rl}
lane, cliff & collins, francis \\
marston, hilary & collins, francis \\
corey, larry & collins, francis \\
collins, francis & erbelding, emily \\
routh, jennifer & myles, renate \\
billet, courtney & myles, renate \\
lane, cliff & myles, renate \\
stover, kathy & myles, renate \\
conrad, patricia & kadlec, robert \\
lane, cliff & grigsby, garrett \\
marston, hilary & grigsby, garrett \\
myles, renate & boyse, natalie \\
kadlec, robert & hassell, david \\
collins, francis & bertuzzi, stefano \\
burklow, john & kaplan, sheila \\
    \end{tabularx}%
    \end{minipage}
    
    \begin{minipage}{\textwidth}
    \caption{The minimum cut that separates Francis Collins (head of the NIH) from Patricia Conrad (Fauci's key assistant) in a sender-receiver network with Fauci removed. Blue edges are cut in the solution and purple nodes are on the Collins side whereas light red nodes are on the Conrad side. The left layout is a force directed layout of the network whereas the right layout is a force directed layout designed to highlight groups in the optimal modularity partition of the network. Many of the cut edges are between nodes with high centrality values (\texttt{routh, kadlec, billet} are in the top 10 PageRank nodes on this graph). }
    \label{fig:stcut-conrad-collins}
    \end{minipage}

%\begin{figure}
    %\centering
    \(\vcenter{\hbox{\includegraphics[width=0.35\linewidth]{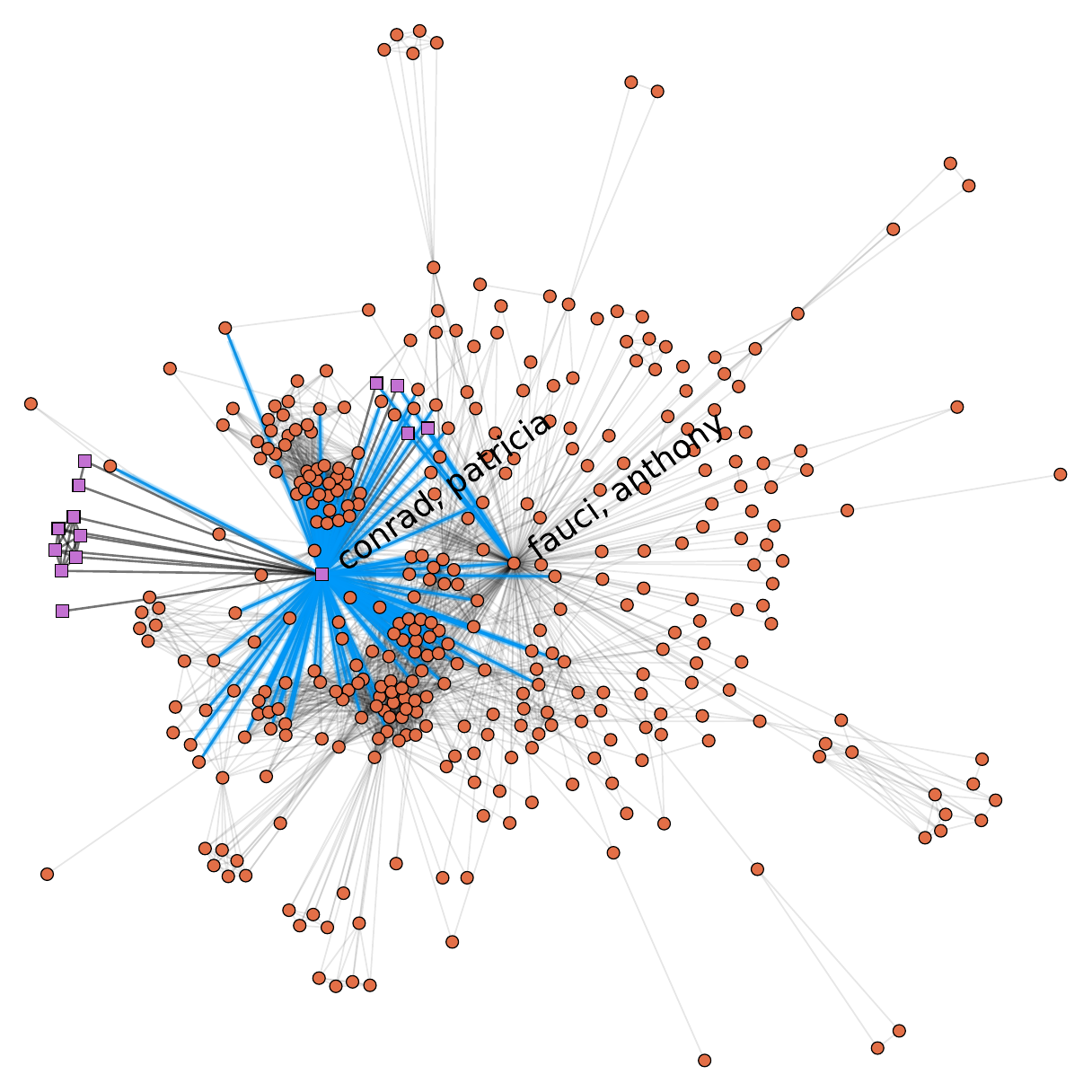}}}\)%
    \(\vcenter{\hbox{\includegraphics[width=0.35\linewidth]{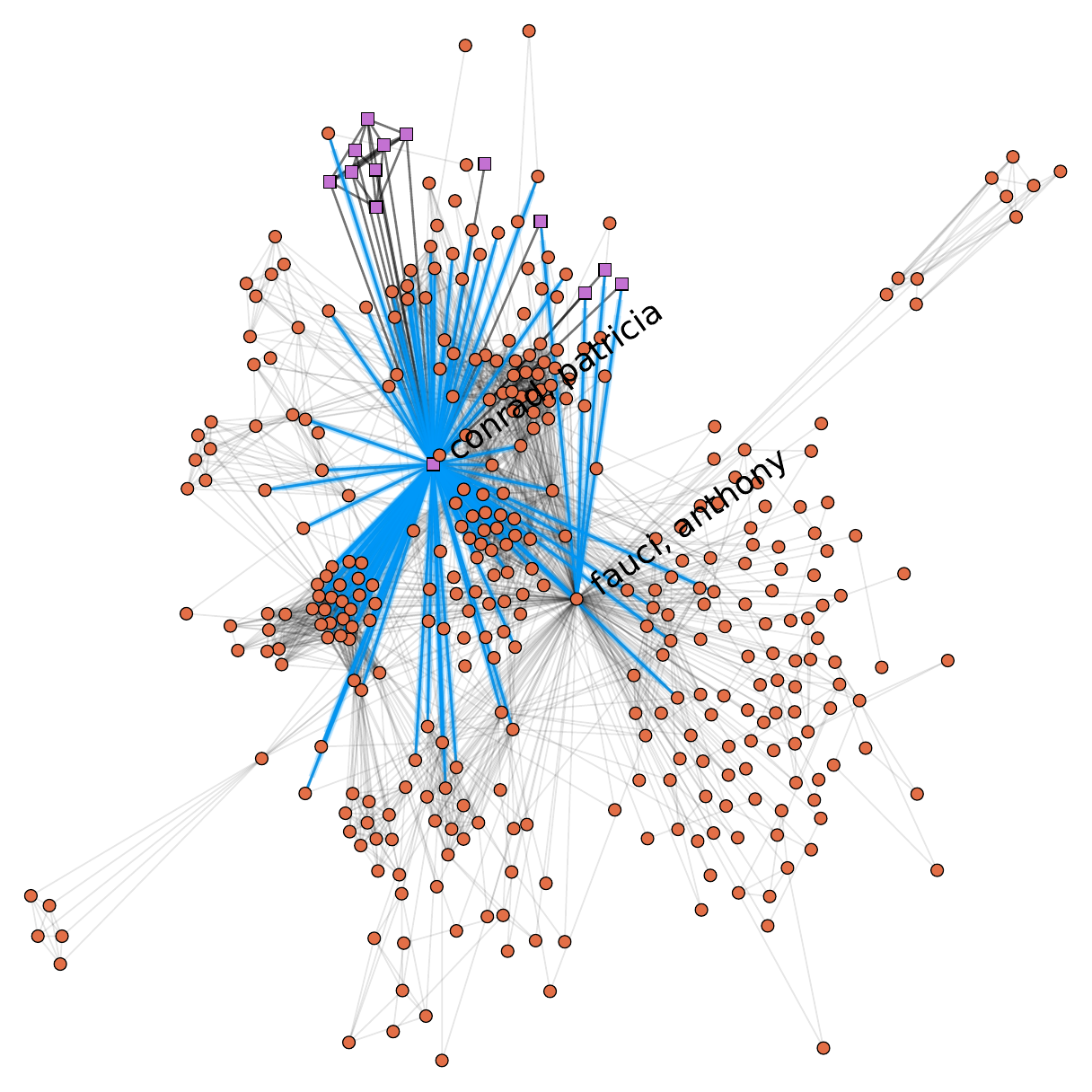}}}\)%
    \begin{minipage}[m]{0.3\linewidth}%
    \centering
    \emph{Conrad set}
    \medskip
    
    \footnotesize%
    \begin{tabular}{c}
goldner, shannah \\
hynds, joanna \\
figliola, mike \\
edwards, sara l \\
mcguffee, tyler ann \\
good-cohn, meredith \\
gathers, shirley \\
amerau, colin c \\
rom, colin \\
baier, bret \\
koerber, ashley \\
griffin, janelle \\
robinson, sae \\
    \end{tabular}%
    \end{minipage}
    
    \begin{minipage}{\textwidth}
    \caption{The minimum cut that separates Anthony Fauci from Patricia Conrad (Fauci's key assistant) in a hypergraph projected to a graph via clique expansion. Blue edges are cut in the solution and purple nodes are on the Conrad side whereas light red nodes are on the Fauci side. The left layout is a force directed layout of the network whereas the right layout is a force directed layout designed to highlight groups in the optimal modularity partition of the network. The nodes on the Conrad side of the cut largely deal with media inquiries and scheduling. }
    \label{fig:stcut-conrad-fauci}
    \end{minipage}
%\end{figure}
    
\end{fullwidthfigure}

Indeed, handling media queries, scheduling, etc.~reveals itself in another minimum cut. In the next example, we consider a weighted hypergraph projection (\texttt{hypergraph-projection} without CC as described below). Each email gives a single hyperedge among all senders and recipients, which is projected onto a clique. Edge weights are the number of emails on that edge. We further filter by removing nodes with only a single edge (ignoring weights). In contrast, we leave Fauci in this network and compute a cut between Fauci and his assistant Patricia Conrad (\figref{fig:stcut-conrad-fauci}). This gives a set of 9 nodes involving media inquires, including a Fox News anchor's repeated requests to interview Fauci. 

Overall, minimum cut analysis is effective at finding meaningful partitions of this network and has the advantage of being a simple method.

%There have been tremendous development in social network analysis methodology since this point. 

%The graphs, networks, and hypergraphs that result from these data are small compared with the size of many modern datasets, yet they are not so small as to permit trivial analysis. 

\subsection{Conductance and minimum normalized cut}
The graphs derived from the data are small enough to allow us to use combinatorial optimization techniques to solve classically hard problems optimally -- that is, we need not use heuristics or approximations to study solutions. One surprising result here was a different between the optimal \emph{normalized cut} set and the optimal \emph{minimum conductance} set. These measures are closely related and frequently interchanged when used in algorithms. This causes a perception that the sets identified by normalized cut optimization and conductance optimization should be similar. Here, we show a natural example where the results sets are extremely different (\figref{fig:cond-vs-ncut}). These use the simple, unweighted \texttt{tofrom-nofauci} graph from below with CC lists included and without Fauci. The optimal normalized cut set is a small group involved in setting up an interview for Fauci -- another \emph{media interaction set}. The optimal conductance set is a large group centered around Collins and other NIH groups. This example serves as a useful reminder that the precise details of the objective functions matter when applied to a specific dataset.

\paragraph{Details and Methods}
For the purposes of being precise, let $G=(V,E)$ be an undirected, unweighted, and connected simple graph. The normalized cut, $\ncut$, of a set $S$ is 
\[ \ncut(S) = \frac{\cut(S)}{\vol(S)} + \frac{\cut(\Sbar)}{\vol(\Sbar)}, \] 
where $\cut(S)$ is the sum of weights of edges cut,  $\vol(S)$ is the sum of weighted degrees of vertices in $S$, and $\Sbar$ is the complement set of vertices $\Sbar = V \ S$. 
In comparison, the conductance, $\phi$, of a set $S$ is 
\[ \phi(S) = \frac{\cut(S)}{\min(\vol(S),\vol(\Sbar))}. \]
Although these two measures are different, we have 
\[ \cond(S) \le \ncut(S) \le 2 \cond(S), \] 
\[ \ncut(S)/2 \le \cond(S) \le \ncut(S). \] 
Both measures $\cond(S)$ and $\ncut(S)$ are NP-hard to optimize in general. Consequently methods designed for approximating conductance often implicitly or explicitly solve problems for $\ncut$ instead of $\cond$ --- the two measures only differ by a factor of two after all. 

Given the weighted graph loaded from data, we remove self-loop entries and edge weights to get the simple,  undirected, unweighted network. (The result does not appear using the weights.) To find each solution set, we solve the combinatorial problem using Gurobi's mixed integer linear programming software. This terminates in a few seconds to minutes. The sets identified by the method are shown in \figref{fig:cond-vs-ncut}; this uses the modularity layout of the network. (See details in Appendix.) For comparison, we also show the $s-t$ cut from Collins to Conrad, which identifies a group around Collins in this unweighted graph. (See \secref{sec:stcut-analysis} for more discussion of st-cuts and how we get a bigger partition in the weighted graph.)

\begin{tuftefigure}[t]
\centering
    \includegraphics[width=0.75\linewidth]{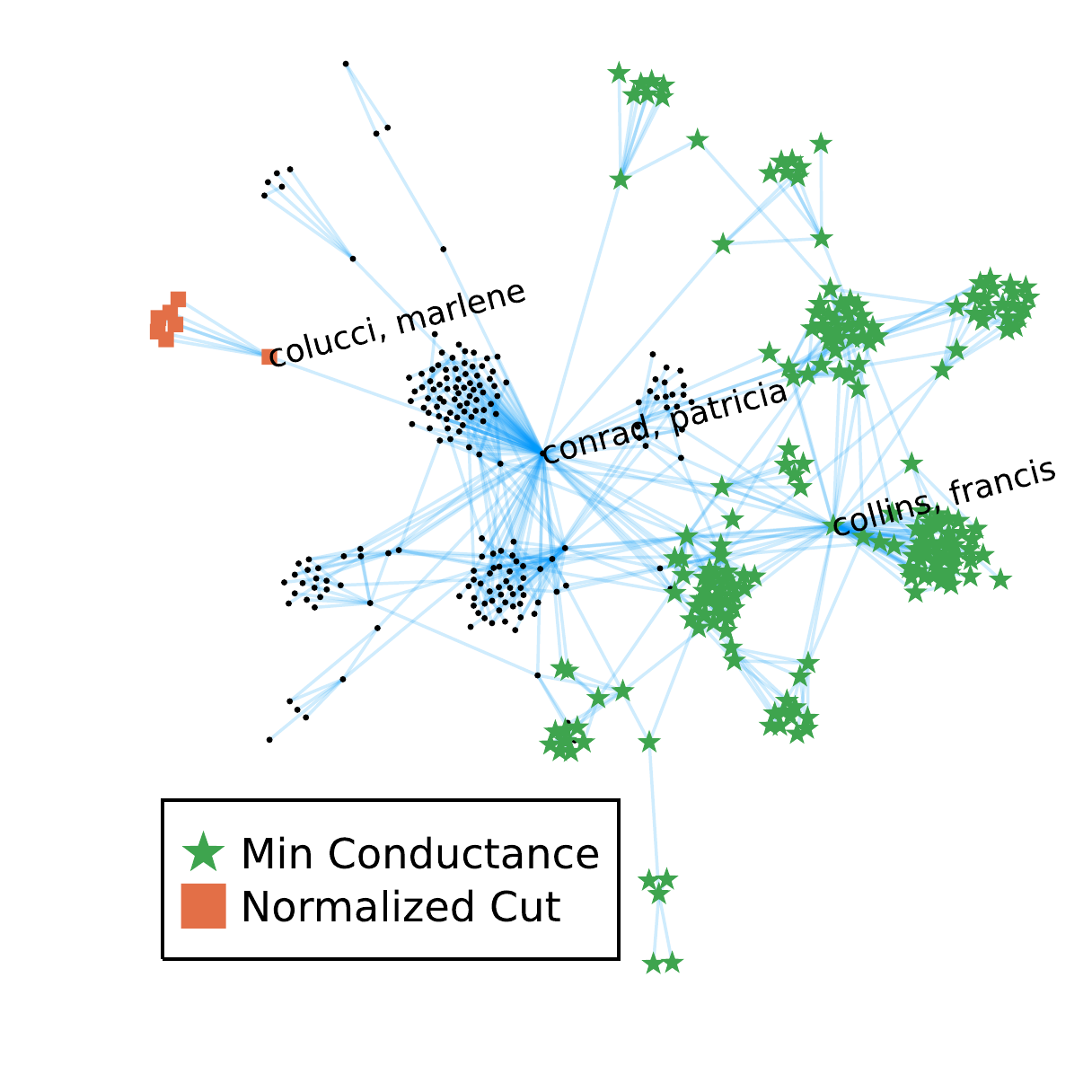} 
    \vspace{-2\baselineskip}
    
    \includegraphics[width=0.375\linewidth]{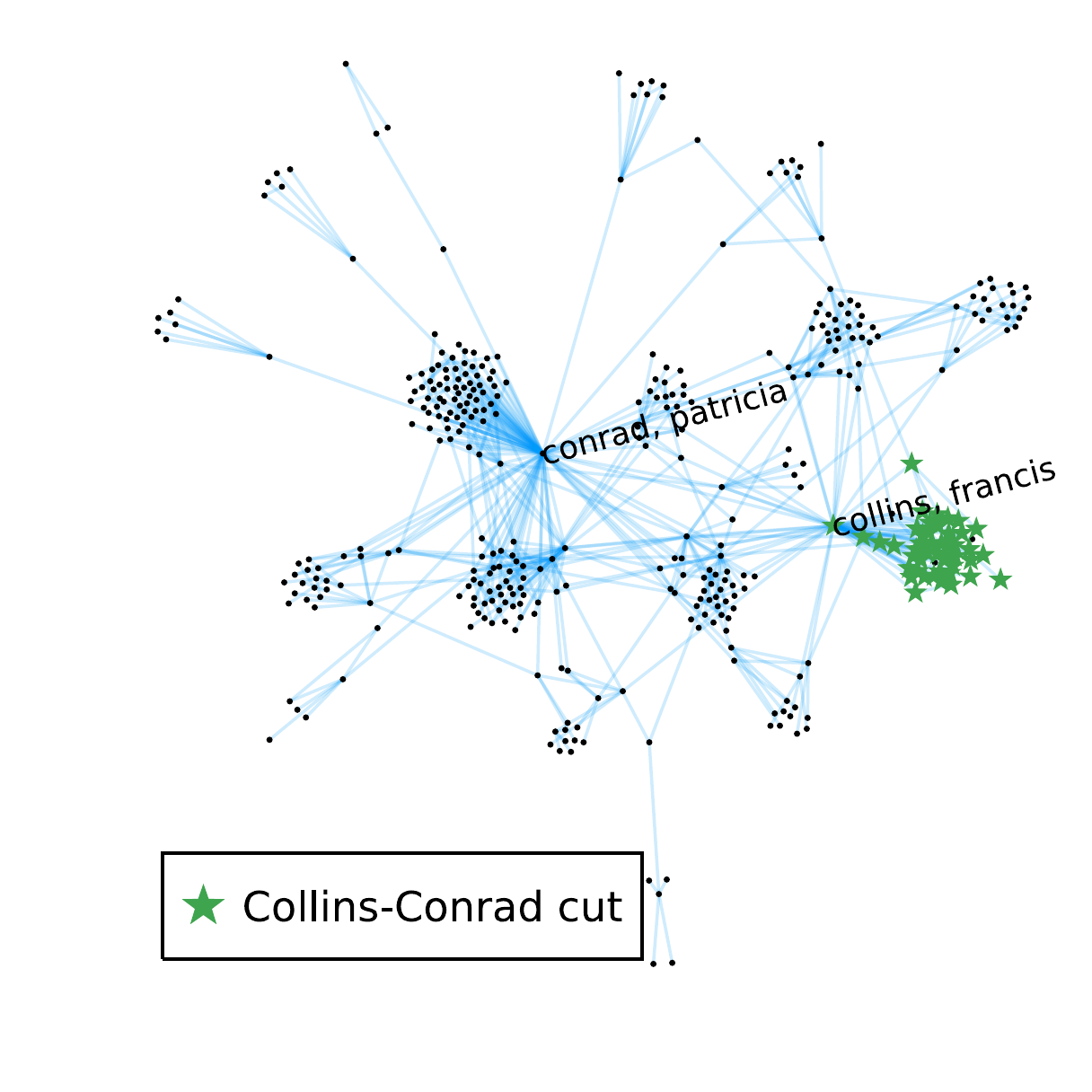}%
    \includegraphics[width=0.375\linewidth]{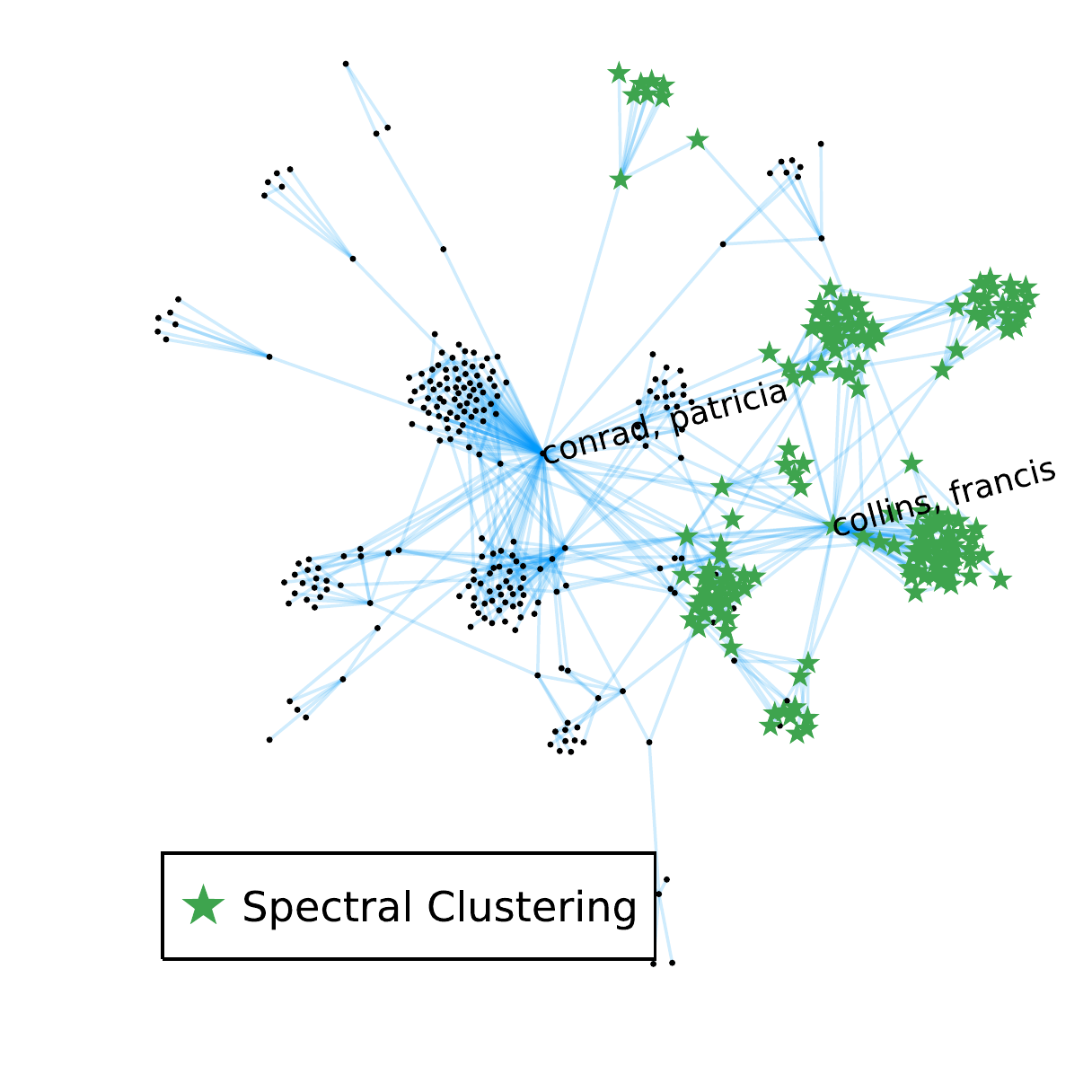}
    
    \footnotesize 
    \begin{tabularx}{0.8\linewidth}{lXXXXX}
    \toprule 
    set & size & cut & vol & $\phi$ & ncut \\
    \midrule 
    conductance & 194 & 35 & 561 & 0.06239 & 0.1193 \\
normalized cut & 7 & 1 & 13 & 0.07692 & 0.07778 \\
Collins-Conrad $s,t$-cut & 45 & 28 & 158 & 0.17722 & 0.20472 \\
spectral cut & 158 & 35 & 479 & 0.07307 & 0.12328 \\
\bottomrule
\end{tabularx}

    \caption{The optimal conductance and normalized cut sets from the undirected, unweighted \texttt{tofrom-cc} graph with 386 nodes and 588 undirected edges are extremely different. The minimum conductance set is about half the graph whereas the minimum normalized cut set is only 7 vertices. The graph layout is computed by emphasizing the groups in a optimal modularity partition of the network. We also show some other simple partitions of the network based on the $s,t$-cut between Collins and Conrad, and also the spectral partition based on a sweepcut of spectral partitioning eigenvector. }
    \label{fig:cond-vs-ncut}
\end{tuftefigure}

%This terminates in a few seconds for minimum conductance and a few minutes for our simplistic way of solving minimum normalized cut.\footnote{For our own convinencem We use an equivalence with the LambdaCC framework~\cite{XXX} and previously written software to exactly solve LambdaCC and modularity problems.) 

\subsection{Modularity partitions}
\label{sec:modularity}
Since we are able to solve many of the combinatorial objectives on this network exactly, for the networks of senders and receivers excluding Fauci (\texttt{tofrom-nofauci-nocc} as a simple graph, we find that the optimal modularity clusters~\cite{Newman2004-community} are characterized by nodes of high betweenness centrality~\cite{Freeman-1977-betweenness,Csardi-2006-igraph} that identify functions and groups in the emails.

See \figref{fig:modularity}, where we label nodes with high betweenness centrality. 

Note the partitioning of agency heads (Collins, Redfield) and task coordinators (Birx, Farrar) as high betweenness nodes in distinct clusters. The clusters identified revolve around different agencies (NIH, CDC, WHO) or functional tasks (handling media requests, budgets), or involve email exchanges around a specific topic, for instance an editorial for the New England Journal of Medicine. %For instance, Routh is involved in many communications and media. Many of the groups have to do with responding to comments from reporters.
Remember that Fauci is involved in almost all of the emails, so the interaction between Redfield, Collins, and Farrar is really modulated by Fauci as well, despite the appearance in this network otherwise. 

Overall, this shows the power of this type of analysis to identify relevant structure in these networks with only a little information. In these networks, the optimal modularity partitions feature nodes with large betweenness centrality, showing another perspective on how this network appears to be constructed with local leaders as one might expect in a working hierarchy. See \tabref{tab:bc}. In that table, we further compare the networks by including the CC labels. 

\marginnote[-20ex]{%
\textbf{More information on people}
\begin{description}
\item[Abutaleb, Yasmeen] is a reporter for the Washington Post
\item[Corey, Larry] was the lead writer for an New England Journal of editorial about COVID Vaccines.
\item [Beigel, John] is the associate director of clinical research.
\item[Farrar, Jeremy] communicated and coordinated with Fauci frequently in terms of global interactions through the WHO Global Preparedness Monitoring Board (GPMB).
\item[Kadlac, Robert] (See above).
\item[Awwad, David] is the NIAID IT field manager~\cite{awwad}.
\item[NIAID OD AM] a mailing list that is frequently forwarded emails for discussion. 
\item[Myles, Renate] is the deputy director for public affairs in the office of communication and public liaison.
\item[Lane, Cliff] a clinical director at NIAID.
\item[Billet, Courtney] was often CCed as a point of coordination for Fauci replying to reporters.
\item[Cabezas, Miriam] helped coordinate emergency budget requests for NIH. 
\end{description}
}

\begin{tuftefigure}[t]
    \centering
    \includegraphics[width=0.75\linewidth]{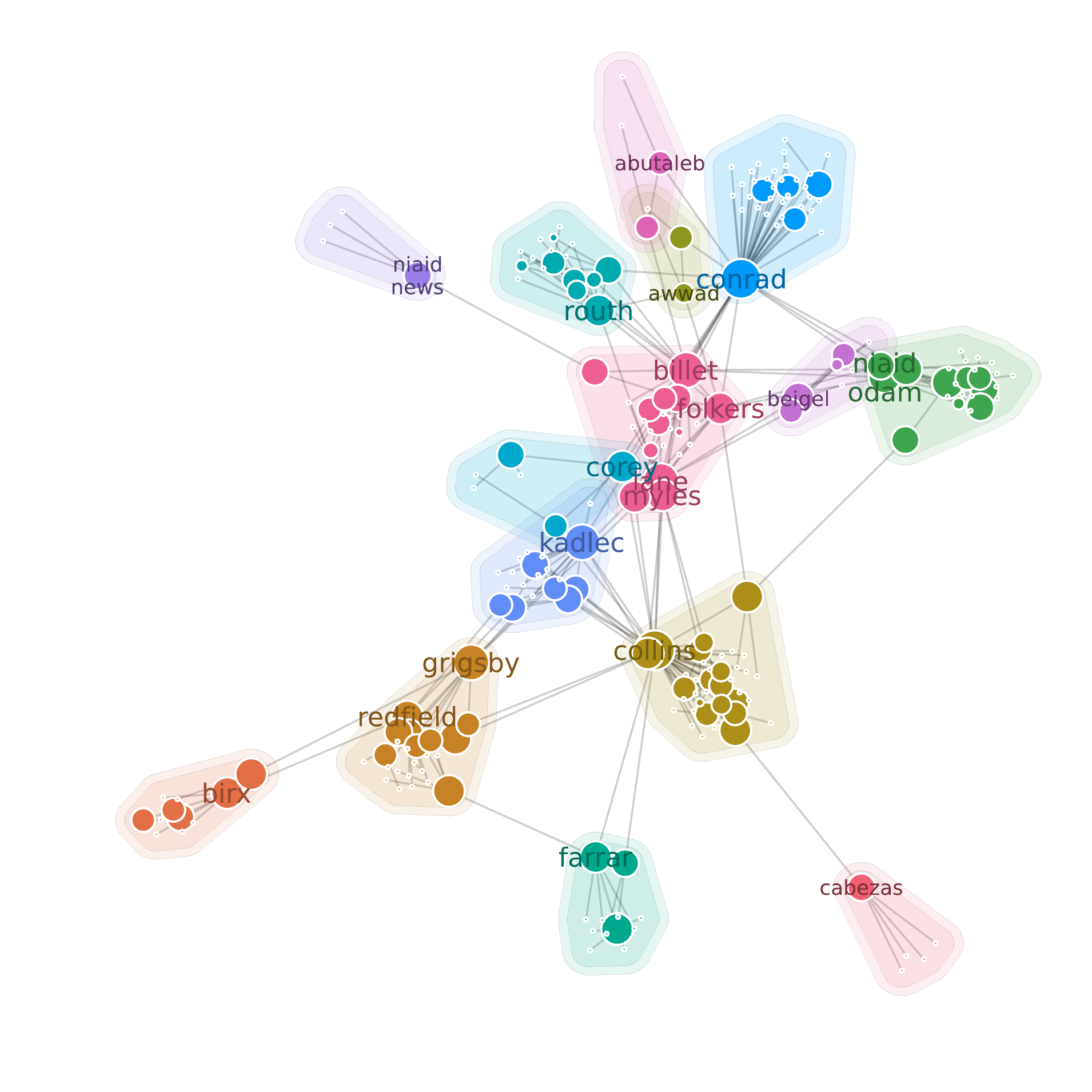}
    \caption{The optimal modularity partition of the network of senders and receivers alone (without Fauci) and reduced to a simple graph are indicated by the colored regions. There are 15 groups and the layout is designed to highlight the modularity groups (see Appendix). We show the 14 most central nodes by betweenness centrality scores in a large fontsize, which labels at least one vertex in all but 5 groups. The small fontsize labels on Abutaleb (rank 46), Awwad (rank 76), Beigel (rank 24), Cabezas (rank 28), and niaid news (rank 33) show key  nodes in clusters that were not top 14 betweenness. Note that many of the agency heads and task leads are identified as key nodes in these networks (Collins, Redfield, Birx, Farrar).  }
    \label{fig:modularity}
\end{tuftefigure}

\begin{table}[p]
\begin{fullwidth}
\includegraphics[width=0.5\linewidth]{figures/modularity-tofrom}%
\includegraphics[width=0.5\linewidth]{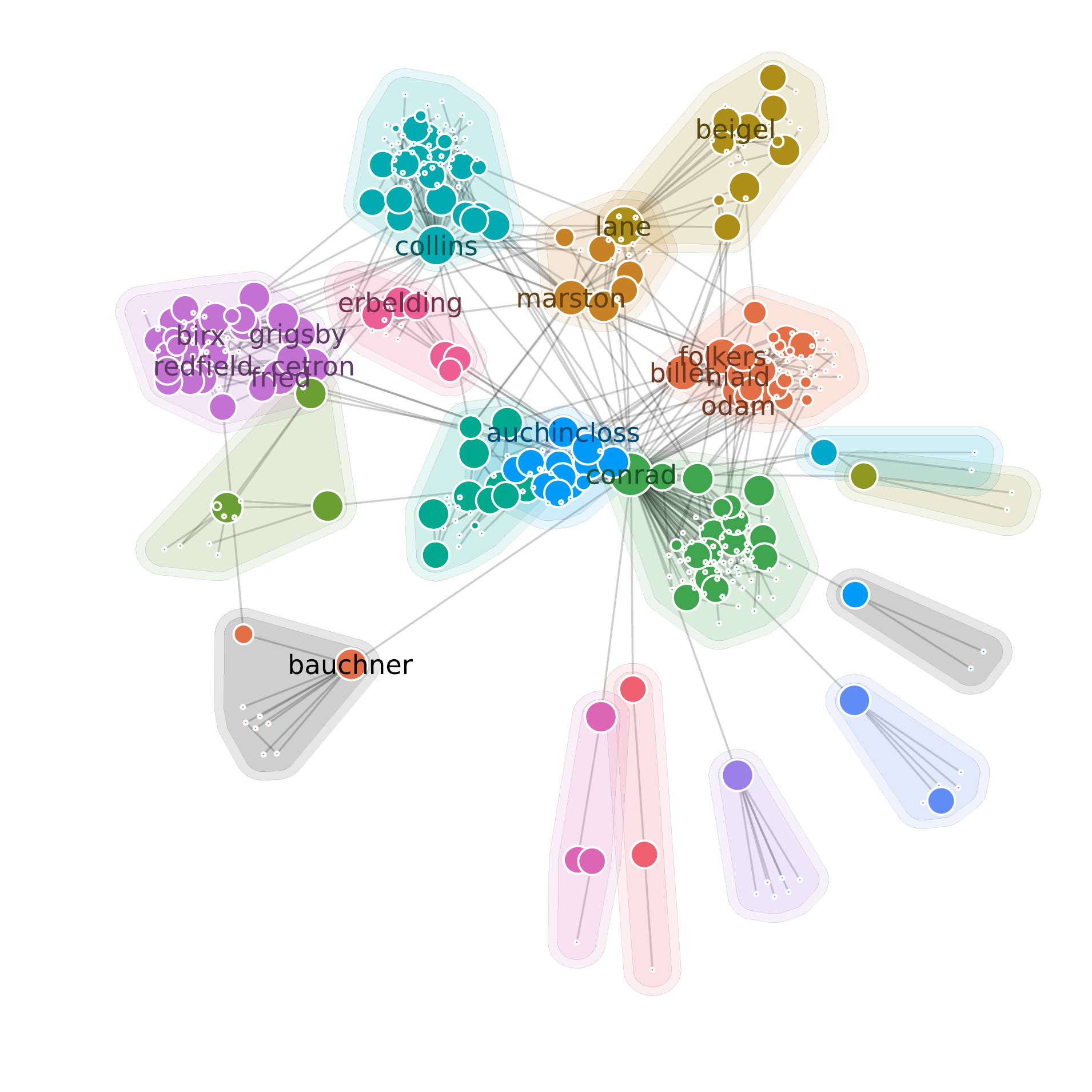}

\centering\fontsize{9}{10}\selectfont
% -- fauci-email-tofrom-5.json
%
\begin{tabular}{*{2}{p{16pt}@{}}p{112pt}}
\toprule
\multicolumn{3}{c}{\texttt{tofrom} without CC} \\
\midrule
\textcolor{LightGray}{1} & 2 & \textcolor[rgb]{0.51,0.42,0.07}{collins, francis, 5} \\
\textcolor{LightGray}{2} & 1 & \textcolor[rgb]{0.0,0.45,0.73}{conrad, patricia, 1} \\
\textcolor{LightGray}{3} & 17 & \textcolor[rgb]{0.29,0.41,0.72}{kadlec, robert, 13} \\
\textcolor{LightGray}{4} & 4 & \textcolor[rgb]{0.7,0.28,0.43}{billet, courtney, 7} \\
\textcolor{LightGray}{5} & 7 & \textcolor[rgb]{0.58,0.38,0.11}{grigsby, garrett, 8} \\
\textcolor{LightGray}{6} & 3 & \textcolor[rgb]{0.7,0.28,0.43}{lane, cliff, 7} \\
\textcolor{LightGray}{7} & 6 & \textcolor[rgb]{0.7,0.28,0.43}{folkers, greg, 7} \\
\textcolor{LightGray}{8} & 38 & \textcolor[rgb]{0.7,0.28,0.43}{myles, renate, 7} \\
\textcolor{LightGray}{9} & 26 & \textcolor[rgb]{0.0,0.5,0.51}{routh, jennifer, 6} \\
\textcolor{LightGray}{10} & 53 & \textcolor[rgb]{0.0,0.5,0.41}{farrar, jeremey, 9} \\
\textcolor{LightGray}{11} & 55 & \textcolor[rgb]{0.0,0.5,0.6}{corey, larry, 11} \\
\textcolor{LightGray}{12} & 16 & \textcolor[rgb]{0.58,0.38,0.11}{redfield, robert, 8} \\
\textcolor{LightGray}{13} & 15 & \textcolor[rgb]{0.18,0.48,0.23}{niaid odam, 3} \\
\textcolor{LightGray}{14} & 10 & \textcolor[rgb]{0.67,0.33,0.21}{birx, deborah, 2} \\
\textcolor{LightGray}{15} & 5 & \textcolor[rgb]{0.7,0.28,0.43}{marston, hilary, 7} \\
\textcolor{LightGray}{16} & 36 & \textcolor[rgb]{0.67,0.33,0.21}{adams, jerome, 2} \\
\textcolor{LightGray}{17} & 37 & \textcolor[rgb]{0.51,0.42,0.07}{tabak, lawrence, 5} \\
\textcolor{LightGray}{18} & 31 & \textcolor[rgb]{0.18,0.48,0.23}{harris, kara, 3} \\
\textcolor{LightGray}{19} & 13 & \textcolor[rgb]{0.18,0.48,0.23}{auchincloss, hugh, 3} \\
\textcolor{LightGray}{20} & 28 & \textcolor[rgb]{0.0,0.5,0.41}{rioux, amelie, 9} \\
\textcolor{LightGray}{21} & 138 & \textcolor[rgb]{0.51,0.42,0.07}{shapiro, neil, 5} \\
\textcolor{LightGray}{22} & 14 & \textcolor[rgb]{0.51,0.42,0.07}{erbelding, emily, 5} \\
\textcolor{LightGray}{23} & 108 & \textcolor[rgb]{0.58,0.38,0.11}{michael, ryan, 8} \\
\textcolor{LightGray}{24} & 12 & \textcolor[rgb]{0.57,0.33,0.62}{beigel, john, 4} \\
\textcolor{LightGray}{25} & 125 & \textcolor[rgb]{0.58,0.38,0.11}{stecker, judy, 8} \\
\bottomrule
\end{tabular}
%
% -- fauci-email-tofrom-cc-5.json
%
\begin{tabular}{*{2}{p{16pt}@{}}p{112pt}}
\toprule
\multicolumn{3}{c}{\texttt{tofrom} with CC} \\
\midrule
2 & \textcolor{LightGray}{1} & \textcolor[rgb]{0.18,0.48,0.23}{conrad, patricia, 3} \\
1 & \textcolor{LightGray}{2} & \textcolor[rgb]{0.7,0.28,0.43}{collins, francis, 7} \\
6 & \textcolor{LightGray}{3} & \textcolor[rgb]{0.0,0.5,0.51}{lane, cliff, 6} \\
4 & \textcolor{LightGray}{4} & \textcolor[rgb]{0.67,0.33,0.21}{billet, courtney, 2} \\
15 & \textcolor{LightGray}{5} & \textcolor[rgb]{0.0,0.5,0.51}{marston, hilary, 6} \\
7 & \textcolor{LightGray}{6} & \textcolor[rgb]{0.67,0.33,0.21}{folkers, greg, 2} \\
5 & \textcolor{LightGray}{7} & \textcolor[rgb]{0.0,0.5,0.41}{grigsby, garrett, 9} \\
210 & \textcolor{LightGray}{8} & \textcolor[rgb]{0.57,0.33,0.62}{fried, linda p, 4} \\
69 & \textcolor{LightGray}{9} & \textcolor[rgb]{0.0,0.5,0.41}{cetron, marty, 9} \\
14 & \textcolor{LightGray}{10} & \textcolor[rgb]{0.57,0.33,0.62}{birx, deborah, 4} \\
-- & \textcolor{LightGray}{11} & \textcolor[rgb]{0.0,0.0,0.0}{bauchner, howard, 18} \\
24 & \textcolor{LightGray}{12} & \textcolor[rgb]{0.0,0.5,0.51}{beigel, john, 6} \\
19 & \textcolor{LightGray}{13} & \textcolor[rgb]{0.0,0.45,0.73}{auchincloss, hugh, 1} \\
22 & \textcolor{LightGray}{14} & \textcolor[rgb]{0.58,0.38,0.11}{erbelding, emily, 8} \\
13 & \textcolor{LightGray}{15} & \textcolor[rgb]{0.67,0.33,0.21}{niaid odam, 2} \\
12 & \textcolor{LightGray}{16} & \textcolor[rgb]{0.57,0.33,0.62}{redfield, robert, 4} \\
3 & \textcolor{LightGray}{17} & \textcolor[rgb]{0.0,0.5,0.41}{kadlec, robert, 9} \\
158 & \textcolor{LightGray}{18} & \textcolor[rgb]{0.57,0.33,0.62}{hahn, stephen, 4} \\
131 & \textcolor{LightGray}{19} & \textcolor[rgb]{0.71,0.28,0.34}{colucci, marlene, 14} \\
88 & \textcolor{LightGray}{20} & \textcolor[rgb]{0.46,0.37,0.68}{schwetz, tara, 12} \\
-- & \textcolor{LightGray}{21} & \textcolor[rgb]{0.0,0.0,0.0}{morrison, stephen, 17} \\
75 & \textcolor{LightGray}{22} & \textcolor[rgb]{0.0,0.5,0.6}{burklow, john, 11} \\
-- & \textcolor{LightGray}{23} & \textcolor[rgb]{0.0,0.0,0.0}{kilmarx, peter, 17} \\
-- & \textcolor{LightGray}{24} & \textcolor[rgb]{0.65,0.3,0.54}{jacobsen, donna, 15} \\
156 & \textcolor{LightGray}{25} & \textcolor[rgb]{0.29,0.41,0.72}{masur, henry, 13} \\
\bottomrule
\end{tabular}

\begin{minipage}{\textwidth}
\caption{For the network of senders and receivers (with Fauci removed) and the same data with the addition of edges due to CC edges (see the networks described in \secref{sec:graphs}) we see different but related partitions. The table ranks nodes by betweenness centrality and also shows the clusters they are in. The top 25 nodes by betweenness have nodes from all clusters except (at left) 10, 12, 14, 15 (out of 15),   and at right except 5, 10, 16 (out of 18). Note how Cliff Lane has a different role when adding in CC-edges and moves to a role between various clinical (Beigel) and communication groups (Marston). }
\label{tab:bc}
\end{minipage}
\end{fullwidth}
\end{table}

\section{Summary of raw data}
\label{sec:raw}
Jason Leopold submitted a freedom of information act request to obtain email surrounding the initial response of United States federal agencies including the National Institutes of Health (NIH) and Centers for Disease Control (CDC) regarding the COVID-19 pandemic.  The result was a 3234 page PDF document~\cite{Leopold-2021-fauci-emails} consisting of emails that Anthony Fauci, the head of the national institute of allergy and infectious disease (NIAID), send between approximately February 2020 and April 2020. Consequently, to be included in the data, the information must have been included in an email that Fauci sent. 

Many email clients include ``reply data'' in the email information, consequently, we are able to infer some amount of communication outside of only what Fauci sent. For example, consider the email in \figref{fig:page-1}. This shows a reply from Fauci to another group with a long CC list. This is in response to a previous email from the same group. 

\begin{tuftefigure}
    \centering
    \includegraphics[width=\linewidth]{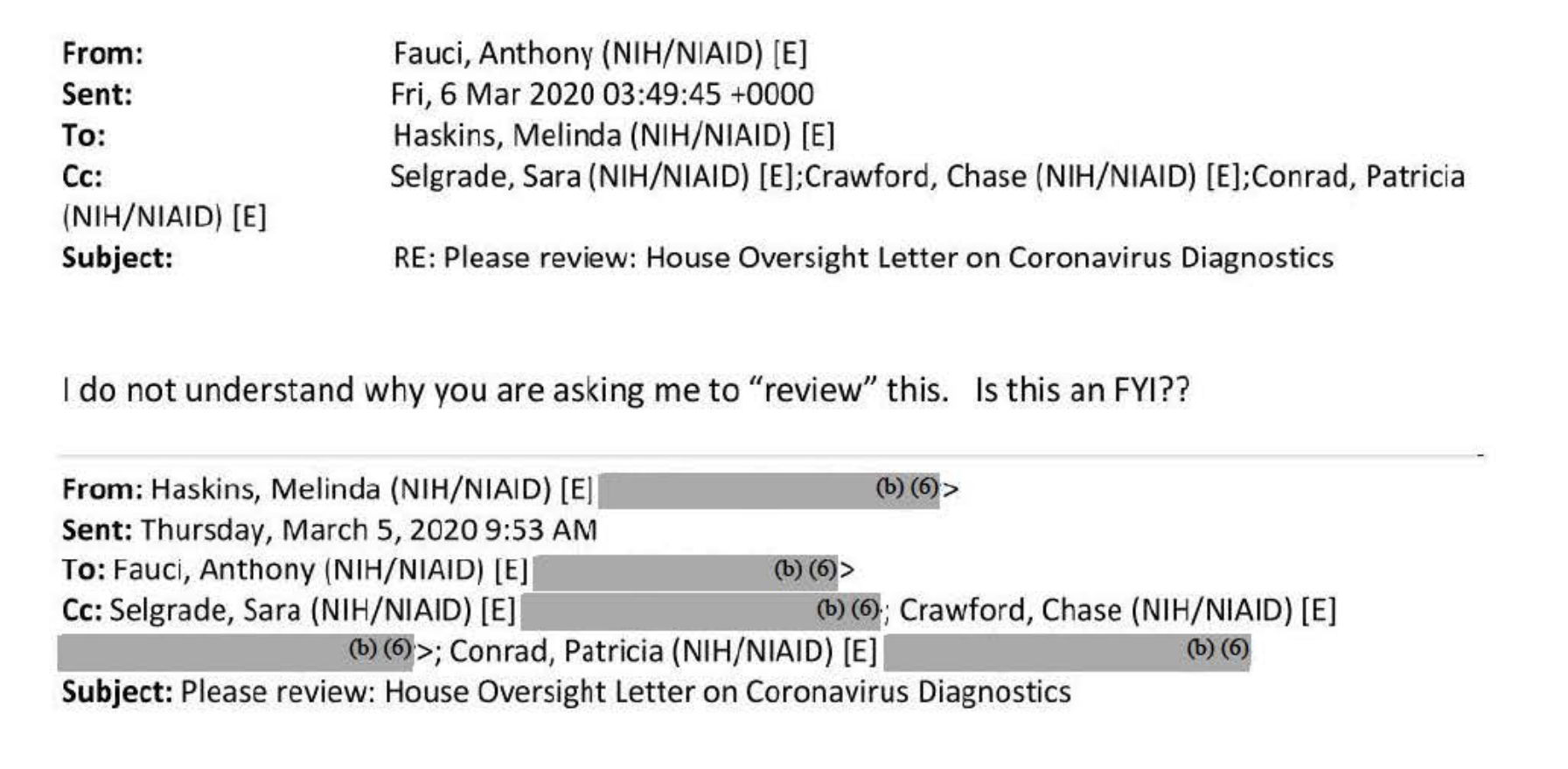}
    \caption{The first page from the PDF file released as part of the freedom of information act request regarding Fauci's email contains the entirety of Fauci's sent email including information (partially redacted) on the email Fauci was replying to. From this page, we are able to extract information on two emails: (i) an email from Fauci to Haskins with a CC to Selgrade, Crawford, and Conrad on 2020-03-06 and (ii) an email from Haskins to Fauci with a CC to Selgrade, Crawford, and Conrad on 2020-03-05. While we have the text of Fauci's email, the text of the original email is redacted.  }
    \label{fig:page-1}
\end{tuftefigure}

\section{Summary of processed data}
\label{sec:processed}

% Austin - Description of automated parsing

The PDF was converted to text and then formatted into a \texttt{json} digest.
The final digest contains
2,761 emails among
1,303 individuals in
1,289 email threads.

The PDF was first converted to a text file with the \texttt{pdftotext} program.\footnote{Specifically, the command \texttt{pdftotext -layout -r 300 leopold-nih-foia-anthony-fauci-emails.pdf}}
New emails in the text begin with a \emph{from line} containing ``From:'', as in \figref{fig:page-1}, to identify the sender of the email.
The file was segmented into chunks of text corresponding to email threads;
the start of a thread was considered to be a from line with Fauci as sender that also began with a form feed character (indicating a new page of the pdf).
The emails within a thread were found by from lines.

The start of the emails contained clear delimiters for the sender, timestamp, recipient list, cc list, and subject (\figref{fig:page-1}). The body of the email was then taken to be all text after the subject and before the next email in the thread.

Timestamps appeared in ten different formats that could be parsed by Python's \texttt{datetime.strptime} function.
The main challenge was handling the numerous errors in the PDF to text conversion.
For example, ``Thursday'' might appear as ``Thu rsday'' or the number 1 and letter l were often interchanged.
Parsing the timestamp involved several general string substitutions and many manual rules for special cases.
We successfully parsed timestamps for 86.5\% of identified emails, and we omitted emails for which we could not parse a timestamp.

The sender, recipient list, and cc list were handled similarly.
For the recipient and cc lists, individuals were separated by the semicolon `;' (the cc list in \figref{fig:page-1} has two semicolons for the three individuals).
Standardizing names involved both automation and considerable manual inspection.
There were issues with text conversion; for instance, ``fauci'' was parsed into several textual variants, including ``f auci,'' ``f.aucl,'' ``fa uci,'' ``fa11ci,'' and ``fauc i.''
Also, one individual could appear with multiple variants on their name or address.
For example, the individual Cliff Lane appeared as ``Lane, Cliff,'' ``Cliff Lane,'' and ``clane@niaid.nih.gov'' in different emails.
The standardization process was iterative.
Given a tentative list of names, we used matching algorithms to find possible duplicates, and these were often checked by manually inspecting the PDF.
Sometimes, emails were sent on behalf of someone else (e.g., Patricia Conrad on behalf of Anthony Fauci).
We treated these as their own ``names'' rather than attributing to one of the parties.
We omitted any emails where we could not identify a sender or at least one recipient, which occurred in 5.1\% of the cases. 
The omissions were mostly caused by redactions or severe errors in the PDF to text conversion.

We also identified federal organizations to which individuals belonged via designations in the email names (e.g., ``NIH'' appearing after all names in \figref{fig:page-1}).
Organization affiliations were
National Institutes of Health (NIH),
Health and Human Services (HHS),
Centers for Disease Control and Prevention (CDC),
the Food and Drug Administration (FDA),
Office of the Secretary (OS), and
the Executive Office of the President (EOP).
Around 26.6\% of individuals were identified as belonging to one of these organizations, and all of the memberships were manually verified.

%\section{Edge cases} 

\section{Example uses}\label{sec:examples}
The subsequent \texttt{json} files are suitable for many types of studies at the intersection of sociology and network science. We describe a few examples.

\subsection{Network analysis}
\label{sec:graphs}
The data can be modeled in terms of a number of different networks that we describe here. Note that there are many other possible networks. For instance, although Fauci was removed from many of these networks, they all could have Fauci in them too. 

\begin{description}
\item[\texttt{repliedto-nofauci}] This is a weighted network that enumerates \emph{replied-to} relationships. We have an edge from $u$ to $v$ if $u$ replied to $v$'s email and then weight the edge with the largest number of interactions in either direction. We remove Fauci from this view of the network to study the view without his emails. This network is an instance of a temporal motif network using a ``replied-to'' temporal motif~\cite{Paranjape2017}. We then remove everyone outside of the largest connected component. 

\item[\texttt{tofrom-nofauci-nocc}] This is a weighted network that has an edge between the sender and recipients of an email (excluding the CC list), weighted by the largest number of interactions in either direction. In this network, we remove emails with more than 5 recipients to focus on \emph{work} behavior instead of \emph{broadcast} behavior. This omits, for instance, weekly emails that detail spending of newly allocated funds to address the pandemic that were often sent to around 20 individuals. We also remove everyone outside the largest connected component. 

\item[\texttt{tofrom-nofauci}] This is the same network above, but expanded to include the CC lists in the number of recipients. The same limit of 5 recipients applies. 

\item[\texttt{hypergraph-projection-nocc}] This is a weighted network that is a network projection of the email hypergraph where each email indicates a hyperedge among the sender and recipients. We then form the clique projection of the hypergraph, where each hyperedge induces a fully connected set of edges among all participants. The weight on an edge in the network are the number of hyperedges that share that edge. The graph is naturally undirected. Because this omits CC lists from each hyperedge, the graph can easily be disconnected if an email arrived via a CC edge. To focus the data analysis, we remove any individual who has only a single edge in the graph (with any weight).

\item[\texttt{hypergraph-projection}] This version of the network adds CCed recipients to the hyperedge for each email. This remains 
disconnected largely due to email lists and BCC-events in the data (see \figref{fig:disconnected} for an instance of a list on page 128 and page 1508 in the PDF~\citet{Leopold-2021-fauci-emails} for an instance of a BCC) even though Fauci remains in this data. Other disconnections are due to parsing errors. There are 35 nodes that are removed due to disconnection. 

\end{description}

\begin{tuftefigure}[t]
    \centering
    \includegraphics[width=0.75\linewidth]{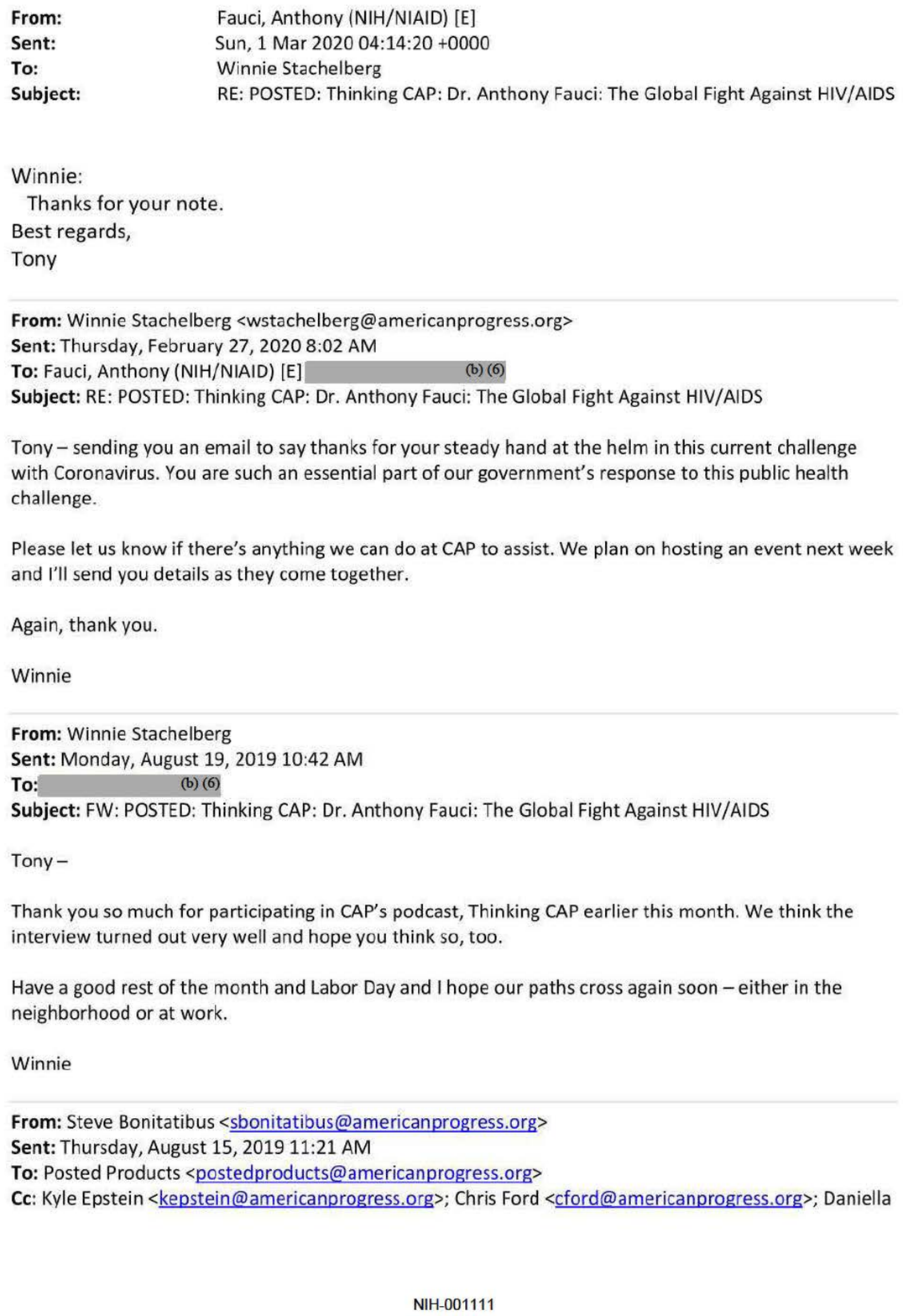}
    \caption{An example email change that produces a disconnected component. In this case, a mailing list ``posted products'' generated an email to multiple people, that were forwarded to Fauci. But Fauci is disconnected  from the original email. This could be addressed by adding links based on the threading, although we did not pursue this avenue in our analysis. }
    \label{fig:disconnected}
\end{tuftefigure}

These are all weighted networks. Consequently, we analyze them as both simple networks (with edge weights and self-loops removed) and the weighted networks depending on the type of analysis. Basic statistics of the networks are given in \tabref{tab:graphs}.

\begin{fullwidthtable}[b]
\footnotesize
    \begin{tabularx}{\linewidth}{@{}>{\raggedright}p{100pt}@{}XXXXXl@{\qquad}XXXXXXl@{}}
    \toprule 
    graph & nodes & \multicolumn{5}{l}{simple graph} & \multicolumn{7}{@{}l}{weighted graph} \\
    \cmidrule(lr){3-7}
    \cmidrule(r){8-14}
    %\cmidrule{7-12}
   & & edges & max deg & mean deg & med deg & $\lambda_2$
     & loops & vol & loop vol & max wdeg & mean wdeg & med wdeg & $\lambda_2$\\
    \midrule 
\texttt{repliedto-nofauci} & 46 & 58 & 18 & 2.5 & 1 & 0.0167 & 2 & 435 & 7 & 91 & 9.5 & 3 & 0.0082\\
\texttt{hypergraph-proj-nocc} & 366 & 2580 & 263 & 14.1 & 6 & 0.0536 & 0 & 13072 & 0 & 1985 & 35.7 & 12 & 0.0346\\
\texttt{hypergraph-proj} & 869 & 7140 & 685 & 16.4 & 7 & 0.0826 & 0 & 76420 & 0 & 4473 & 87.9 & 11 & 0.0254\\
\texttt{tofrom-nofauci-nocc} & 233 & 324 & 44 & 2.8 & 1 & 0.0324 & 2 & 1164 & 2 & 102 & 5.0 & 2 & 0.0305\\
\texttt{tofrom-nofauci} & 386 & 588 & 97 & 3.0 & 2 & 0.0457 & 9 & 2179 & 15 & 248 & 5.6 & 2 & 0.0316\\
\bottomrule 
    \end{tabularx}
    
    \caption{The 5 canonical graphs we derive from the email data along with some simple statistics. Each graph is connected, and there is a simple version without weights and self-loops along with a weighted version that has integer edge weights along with possible self-loops. The number of edges is the count of undirected edges, so there are twice this many non-zeros in the adjacency matrix of the simple graph. The weighted graph also has loops, which gives twice this many non-zeros plus the number of loops in the adjacency matrix. We also show the total volume (sum of weighted degrees) of the weighted graph along with max, median (med), and mean statistics on the degrees of the simple (deg) and weighted graphs (wdeg). Finally, we show the value of $\lambda_2$ associated with the normalized Laplacian matrix. The graph names with \texttt{nofauci} do not include Fauci's node and those with \texttt{nocc} omit the CC lists from the construction whereas those without this treat \emph{CC} lists equivalently with other recipients.}
    \label{tab:graphs}
\end{fullwidthtable}

\paragraph{PageRank and Degree centrality scores}
As an example use case, we can study how centrality changes with graph construction.
PageRank and Degree centrality are two heavily studied centrality measures for graphs. For undirected graphs, such as those we are studying, it is often the case that the two are highly related. For the 5 graphs we construct -- after removing Fauci from each graph -- we find considerable differences in a simple analysis of the rankings. See \hyperref[tab:pagerank]{tables~\ref{tab:pagerank}} \hyperref[tab:degree]{and \ref{tab:degree}} for the 10 rankings by PageRank and degree centrality in each of these 5 weighted, undirected graphs. There are also considerable differences between graphs, showing how each construction highlights different features of the network.

\begin{table}[p]
\begin{fullwidth}
\centering 
%
% -- fauci-email-repliedto.json
%
\begin{tabular}{*{5}{p{16pt}@{}}p{126pt}}
\toprule
\multicolumn{6}{c}{\texttt{repliedto}} \\
\midrule
\textcolor{LightGray}{1} & 2 & 3 & 2 & 2 & collins, francis, 0.171566 \\
\textcolor{LightGray}{2} & 1 & 1 & 1 & 1 & conrad, patricia, 0.095884 \\
\textcolor{LightGray}{3} & 10 & 6 & 6 & 6 & routh, jennifer, 0.066093 \\
\textcolor{LightGray}{4} & 3 & 2 & 3 & 3 & billet, courtney, 0.063542 \\
\textcolor{LightGray}{5} & 105 & 104 & 19 & 10 & goldner, shannah, 0.048336 \\
\textcolor{LightGray}{6} & 11 & 9 & 8 & 5 & tabak, lawrence, 0.044542 \\
\textcolor{LightGray}{7} & 31 & 101 & 7 & 22 & shapiro, neil, 0.036022 \\
\textcolor{LightGray}{8} & 144 & 149 & 36 & 381 & o'malley, devin, 0.024162 \\
\textcolor{LightGray}{9} & -- & 444 & -- & 83 & antoniak, cynthia, 0.021884 \\
\textcolor{LightGray}{10} & 6 & 13 & 9 & 21 & kadlec, robert, 0.020327 \\
\bottomrule
\end{tabular}

%
% -- fauci-email-hypergraph-projection.json
%
\begin{tabular}{*{5}{p{16pt}@{}}p{126pt}}
\toprule
\multicolumn{6}{c}{\texttt{hypergraph-projection} without CC} \\
\midrule
2 & \textcolor{LightGray}{1} & 1 & 1 & 1 & conrad, patricia, 0.028052 \\
1 & \textcolor{LightGray}{2} & 3 & 2 & 2 & collins, francis, 0.025174 \\
4 & \textcolor{LightGray}{3} & 2 & 3 & 3 & billet, courtney, 0.017615 \\
18 & \textcolor{LightGray}{4} & 8 & 16 & 8 & lane, cliff, 0.016952 \\
-- & \textcolor{LightGray}{5} & 10 & 15 & 44 & redfield, robert, 0.013901 \\
10 & \textcolor{LightGray}{6} & 13 & 9 & 21 & kadlec, robert, 0.013061 \\
25 & \textcolor{LightGray}{7} & 4 & 4 & 4 & folkers, greg, 0.013043 \\
-- & \textcolor{LightGray}{8} & 5 & 23 & 7 & marston, hilary, 0.01207 \\
-- & \textcolor{LightGray}{9} & 16 & 18 & 18 & auchincloss, hugh, 0.011343 \\
3 & \textcolor{LightGray}{10} & 6 & 6 & 6 & routh, jennifer, 0.010995 \\
\bottomrule
\end{tabular}
\qquad
%
% -- fauci-email-hypergraph-projection-cc.json
%
\begin{tabular}{*{5}{p{16pt}@{}}p{126pt}}
\toprule
\multicolumn{6}{c}{\texttt{hypergraph-projection} with CC} \\
\midrule
2 & 1 & \textcolor{LightGray}{1} & 1 & 1 & conrad, patricia, 0.045265 \\
4 & 3 & \textcolor{LightGray}{2} & 3 & 3 & billet, courtney, 0.016938 \\
1 & 2 & \textcolor{LightGray}{3} & 2 & 2 & collins, francis, 0.016757 \\
25 & 7 & \textcolor{LightGray}{4} & 4 & 4 & folkers, greg, 0.015869 \\
-- & 8 & \textcolor{LightGray}{5} & 23 & 7 & marston, hilary, 0.015379 \\
3 & 10 & \textcolor{LightGray}{6} & 6 & 6 & routh, jennifer, 0.013172 \\
-- & 52 & \textcolor{LightGray}{7} & 55 & 12 & stover, kathy, 0.011924 \\
18 & 4 & \textcolor{LightGray}{8} & 16 & 8 & lane, cliff, 0.011297 \\
6 & 11 & \textcolor{LightGray}{9} & 8 & 5 & tabak, lawrence, 0.00933 \\
-- & 5 & \textcolor{LightGray}{10} & 15 & 44 & redfield, robert, 0.008778 \\
\bottomrule
\end{tabular}

%
% -- fauci-email-tofrom-5.json
%
\begin{tabular}{*{5}{p{16pt}@{}}p{126pt}}
\toprule
\multicolumn{6}{c}{\texttt{tofrom} without CC} \\
\midrule
2 & 1 & 1 & \textcolor{LightGray}{1} & 1 & conrad, patricia, 0.073189 \\
1 & 2 & 3 & \textcolor{LightGray}{2} & 2 & collins, francis, 0.061331 \\
4 & 3 & 2 & \textcolor{LightGray}{3} & 3 & billet, courtney, 0.023909 \\
25 & 7 & 4 & \textcolor{LightGray}{4} & 4 & folkers, greg, 0.023114 \\
-- & 21 & 42 & \textcolor{LightGray}{5} & 9 & niaid odam, 0.021508 \\
3 & 10 & 6 & \textcolor{LightGray}{6} & 6 & routh, jennifer, 0.021158 \\
7 & 31 & 101 & \textcolor{LightGray}{7} & 22 & shapiro, neil, 0.019681 \\
6 & 11 & 9 & \textcolor{LightGray}{8} & 5 & tabak, lawrence, 0.019056 \\
10 & 6 & 13 & \textcolor{LightGray}{9} & 21 & kadlec, robert, 0.01531 \\
26 & 40 & 36 & \textcolor{LightGray}{10} & 19 & farrar, jeremey, 0.014289 \\
\bottomrule
\end{tabular}
%
% -- fauci-email-tofrom-cc-5.json
%
\qquad
\begin{tabular}{*{5}{p{16pt}@{}}p{126pt}}
\toprule
\multicolumn{6}{c}{\texttt{tofrom} with CC} \\
\midrule
2 & 1 & 1 & 1 & \textcolor{LightGray}{1} & conrad, patricia, 0.086833 \\
1 & 2 & 3 & 2 & \textcolor{LightGray}{2} & collins, francis, 0.040497 \\
4 & 3 & 2 & 3 & \textcolor{LightGray}{3} & billet, courtney, 0.032698 \\
25 & 7 & 4 & 4 & \textcolor{LightGray}{4} & folkers, greg, 0.024633 \\
6 & 11 & 9 & 8 & \textcolor{LightGray}{5} & tabak, lawrence, 0.014353 \\
3 & 10 & 6 & 6 & \textcolor{LightGray}{6} & routh, jennifer, 0.013892 \\
-- & 8 & 5 & 23 & \textcolor{LightGray}{7} & marston, hilary, 0.013344 \\
18 & 4 & 8 & 16 & \textcolor{LightGray}{8} & lane, cliff, 0.012137 \\
-- & 21 & 42 & 5 & \textcolor{LightGray}{9} & niaid odam, 0.011972 \\
5 & 105 & 104 & 19 & \textcolor{LightGray}{10} & goldner, shannah, 0.011434 \\
\bottomrule
\end{tabular}
\caption{PageRank centrality rankings (with $\alpha=0.85$) in 5 different weighted graphs derived from the data. All the graphs are undirected, and Anthony Fauci has been removed from all of these graphs, rendering some of them disconnected. The values prefixing each name are the ranks in alternative orderings. The order of these is the same as the order of tables and the ordered list is shown in light gray to emphasize differences in other lists. For instance,  \texttt{antoniak} is ranked 9 in the \texttt{repliedto} graph but ranked 444 in the hypergraph projection with CC and 83 in the tofrom with CCed nodes. Other individuals of note with large changes in rank include \texttt{stover}, \texttt{redfield}, \texttt{shapiro}, \texttt{farrar}, \texttt{goldner}, \texttt{marston}, \texttt{folkers}.}
\label{tab:pagerank}
\end{fullwidth}
\end{table}

\begin{table}[p]
\begin{fullwidth}
\centering
%
% -- fauci-email-repliedto.json
%
\begin{tabular}{*{5}{p{16pt}@{}}p{112pt}}
\toprule
\multicolumn{6}{c}{\texttt{repliedto}} \\
\midrule
\textcolor{LightGray}{1} & 2 & 1 & 1 & 1 & conrad, patricia, 91.0 \\
\textcolor{LightGray}{2} & 104 & 121 & 13 & 8 & goldner, shannah, 51.0 \\
\textcolor{LightGray}{3} & 1 & 40 & 2 & 3 & collins, francis, 39.0 \\
\textcolor{LightGray}{4} & 12 & 38 & 3 & 5 & routh, jennifer, 28.0 \\
\textcolor{LightGray}{5} & 4 & 22 & 5 & 2 & billet, courtney, 24.0 \\
\textcolor{LightGray}{6} & -- & 259 & -- & 50 & antoniak, cynthia, 22.0 \\
\textcolor{LightGray}{7} & 171 & 366 & 43 & 51 & figliola, mike, 15.0 \\
\textcolor{LightGray}{8} & 14 & 76 & 31 & 20 & awwad, david, 14.0 \\
\textcolor{LightGray}{9} & 169 & 192 & 44 & 34 & chugh, latika, 14.0 \\
\textcolor{LightGray}{10} & -- & -- & 74 & 61 & katz, ruth, 13.0 \\
\bottomrule
\end{tabular}

%
% -- fauci-email-hypergraph-projection.json
%
\begin{tabular}{*{5}{p{16pt}@{}}p{112pt}}
\toprule
\multicolumn{6}{c}{\texttt{hypergraph-projection} without CC} \\
\midrule
3 & \textcolor{LightGray}{1} & 40 & 2 & 3 & collins, francis, 276.0 \\
1 & \textcolor{LightGray}{2} & 1 & 1 & 1 & conrad, patricia, 257.0 \\
20 & \textcolor{LightGray}{3} & 44 & 12 & 13 & lane, cliff, 211.0 \\
5 & \textcolor{LightGray}{4} & 22 & 5 & 2 & billet, courtney, 196.0 \\
21 & \textcolor{LightGray}{5} & 2 & 10 & 27 & kadlec, robert, 170.0 \\
-- & \textcolor{LightGray}{6} & 47 & 20 & 64 & redfield, robert, 166.0 \\
-- & \textcolor{LightGray}{7} & 34 & 16 & 9 & marston, hilary, 154.0 \\
11 & \textcolor{LightGray}{8} & 46 & 8 & 6 & tabak, lawrence, 131.0 \\
19 & \textcolor{LightGray}{9} & 29 & 4 & 4 & folkers, greg, 130.0 \\
-- & \textcolor{LightGray}{10} & 53 & 26 & 29 & auchincloss, hugh, 129.0 \\
\bottomrule
\end{tabular}
%
% -- fauci-email-hypergraph-projection-cc.json
%
\qquad
\begin{tabular}{*{5}{p{16pt}@{}}p{112pt}}
\toprule
\multicolumn{6}{c}{\texttt{hypergraph-projection} with CC} \\
\midrule
1 & 2 & \textcolor{LightGray}{1} & 1 & 1 & conrad, patricia, 2027.0 \\
21 & 5 & \textcolor{LightGray}{2} & 10 & 27 & kadlec, robert, 1335.0 \\
-- & 46 & \textcolor{LightGray}{3} & -- & 194 & disbrow, gary, 1114.0 \\
-- & 49 & \textcolor{LightGray}{4} & 165 & -- & yeskey, kevin, 1078.0 \\
-- & -- & \textcolor{LightGray}{5} & -- & -- & hatchett, richard, 1068.0 \\
-- & -- & \textcolor{LightGray}{6} & -- & -- & mecher, carter, 1060.0 \\
-- & -- & \textcolor{LightGray}{7} & -- & -- & caneva, duane, 1060.0 \\
-- & -- & \textcolor{LightGray}{8} & -- & -- & mcnamara, tracey, 1053.0 \\
-- & -- & \textcolor{LightGray}{9} & -- & -- & eva k lee, 1042.0 \\
-- & -- & \textcolor{LightGray}{10} & -- & -- & lawler, james, 1040.0 \\
\bottomrule
\end{tabular}

%
% -- fauci-email-tofrom-5.json
%
\begin{tabular}{*{5}{p{16pt}@{}}p{112pt}}
\toprule
\multicolumn{6}{c}{\texttt{tofrom} without CC} \\
\midrule
1 & 2 & 1 & \textcolor{LightGray}{1} & 1 & conrad, patricia, 102.0 \\
3 & 1 & 40 & \textcolor{LightGray}{2} & 3 & collins, francis, 91.0 \\
4 & 12 & 38 & \textcolor{LightGray}{3} & 5 & routh, jennifer, 38.0 \\
19 & 9 & 29 & \textcolor{LightGray}{4} & 4 & folkers, greg, 36.0 \\
5 & 4 & 22 & \textcolor{LightGray}{5} & 2 & billet, courtney, 35.0 \\
13 & 50 & 124 & \textcolor{LightGray}{6} & 17 & shapiro, neil, 35.0 \\
-- & 52 & 105 & \textcolor{LightGray}{7} & 7 & niaid odam, 33.0 \\
11 & 8 & 46 & \textcolor{LightGray}{8} & 6 & tabak, lawrence, 33.0 \\
23 & 32 & 52 & \textcolor{LightGray}{9} & 26 & myles, renate, 25.0 \\
21 & 5 & 2 & \textcolor{LightGray}{10} & 27 & kadlec, robert, 18.0 \\
\bottomrule
\end{tabular}
%
% -- fauci-email-tofrom-cc-5.json
%
\qquad
\begin{tabular}{*{5}{p{16pt}@{}}p{112pt}}
\toprule
\multicolumn{6}{c}{\texttt{tofrom} with CC} \\
\midrule
1 & 2 & 1 & 1 & \textcolor{LightGray}{1} & conrad, patricia, 248.0 \\
5 & 4 & 22 & 5 & \textcolor{LightGray}{2} & billet, courtney, 109.0 \\
3 & 1 & 40 & 2 & \textcolor{LightGray}{3} & collins, francis, 106.0 \\
19 & 9 & 29 & 4 & \textcolor{LightGray}{4} & folkers, greg, 85.0 \\
4 & 12 & 38 & 3 & \textcolor{LightGray}{5} & routh, jennifer, 47.0 \\
11 & 8 & 46 & 8 & \textcolor{LightGray}{6} & tabak, lawrence, 40.0 \\
-- & 52 & 105 & 7 & \textcolor{LightGray}{7} & niaid odam, 39.0 \\
2 & 104 & 121 & 13 & \textcolor{LightGray}{8} & goldner, shannah, 39.0 \\
-- & 7 & 34 & 16 & \textcolor{LightGray}{9} & marston, hilary, 38.0 \\
-- & 84 & 41 & 46 & \textcolor{LightGray}{10} & stover, kathy, 37.0 \\
\bottomrule
\end{tabular}
\caption{Degree centrality rankings in 5 different weighted graphs derived from the data. All the graphs are undirected, and Anthony Fauci has been removed from all of these graphs, rendering some of them disconnected. The values prefixing each name are the ranks in alternative orderings. The order of these is the same as the order of tables and the ordered list is shown in light gray to emphasize differences in other lists. Note, for instance, that \texttt{awwad} doesn't appear in any of the top 10 PageRank lists.} 
\label{tab:degree}
\end{fullwidth}
\end{table}

\paragraph{Other uses} These graphs are used in the leading examples above in Section~\ref{sec:findings}.

\subsection{Hypergraph analysis}
\label{sec:hypergraph}
The \texttt{hypergraph-projection} data is one example of a hypergraph analysis (as a projected graph).  We now consider the email data as a hypergraph where each email is a hyperedge among the senders and recipients (excluding the CC entries) -- excluding Fauci. We remove any individual that does not have at at least degree 5 in a clique expansion of the resulting graph. The largest connected component of resulting hypergraph has 233 vertices and 254 hyperedges. 

\paragraph{Differences between local diffusions}
A local diffusion in a graph or hypergraph answers the question: \emph{what else might be related to a given node in a graph or hypergraph}. It's an instance of a relationship-by-transitivity-of-relationships study. 
Local diffusion analysis on hypergraphs have been a recently active area. Here, we show how three closely related ideas around PageRank-like diffusions produce strikingly different results on this hypergraph, which indicates it's a useful tool for followup work on comparisons among the implications of these ideas. 

PageRank-like diffusions are \emph{quadratic} or \emph{smoothed} variations on cut problems for graphs and hypergraphs~\cite{Liu-2021-localhyper}. They can be \emph{seeded} on a single node to generate a ranked list of other nodes based on relationship strength. We do this for a sparse PageRank diffusion on a graph projection of the hypergraph, a direct sparse PageRank diffusion on the hypergraph, and a unregularized PageRank diffusion on the hypergraph. (Sparse PageRank diffusions include regularization extra terms to encourage sparse solutions of the PageRank diffusion equations.) The difference in results is shown in \tabref{tab:pagerank-hypergraph}. There are far more differences than one would expect between these solutions. This indicates an area of further study. It possible simple parameter changes or other tools will show how these are more similar than apparent from this simple experiment. 

%The idea is that PageRank is a regularized smoothing operator based on the Laplacian of a graph -- which can then be extended to various notions of the Laplacian of a hypergraph. Alternatively, it is a minorant of a cut. This means we can 

\begin{fullwidthtable}[t]
%\begin{fullwidth}
\footnotesize
\centering
    %
% -- Graph-ACL
%
\begin{tabular}{*{3}{@{}p{15pt}@{}}p{106pt}@{}}
\toprule
\multicolumn{4}{c}{Sparse Seeded PageRank-Graph} \\
\midrule
\textcolor{LightGray}{1} & 1 & 1 & conrad, patricia, 0.150591 \\
\textcolor{LightGray}{2} & 26 & 3 & billet, courtney, 0.031087 \\
\textcolor{LightGray}{3} & 19 & 2 & folkers, greg, 0.023756 \\
\textcolor{LightGray}{4} & 145 & 4 & collins, francis, 0.019185 \\
\textcolor{LightGray}{5} & 85 & 8 & lane, cliff, 0.017899 \\
\textcolor{LightGray}{6} & 5 & 6 & goldner, shannah, 0.01682 \\
\textcolor{LightGray}{7} & 24 & 113 & brennan, patrick, 0.016653 \\
\textcolor{LightGray}{8} & 38 & 10 & marston, hilary, 0.015882 \\
\textcolor{LightGray}{9} & 20 & 26 & lepore, loretta, 0.01522 \\
\textcolor{LightGray}{10} & 35 & 9 & routh, jennifer, 0.014561 \\
\textcolor{LightGray}{11} & 18 & 46 & bonds, michelle, 0.014441 \\
\textcolor{LightGray}{12} & 23 & 120 & fine, amanda, 0.014379 \\
\textcolor{LightGray}{13} & 101 & 17 & kadlec, robert, 0.01374 \\
\textcolor{LightGray}{14} & 164 & 21 & redfield, robert, 0.01257 \\
\textcolor{LightGray}{15} & 40 & 14 & awwad, david, 0.011282 \\
\bottomrule
\end{tabular} 
%
% -- LHPR
%
\begin{tabular}{*{3}{@{}p{15pt}@{}}p{112pt}@{}}
\toprule
\multicolumn{4}{c}{Sparse Seeded PageRank-HyperGraph} \\
\midrule
1 & \textcolor{LightGray}{1} & 1 & conrad, patricia, 0.415602 \\
66 & \textcolor{LightGray}{2} & 23 & katz, ruth, 0.319543 \\
50 & \textcolor{LightGray}{3} & 18 & hynds, joanna, 0.319543 \\
98 & \textcolor{LightGray}{4} & 33 & koerber, ashley, 0.319543 \\
6 & \textcolor{LightGray}{5} & 6 & goldner, shannah, 0.319543 \\
29 & \textcolor{LightGray}{6} & 12 & figliola, mike, 0.312954 \\
139 & \textcolor{LightGray}{7} & 20 & barasch, kimberly, 0.216517 \\
129 & \textcolor{LightGray}{8} & 99 & amerau, colin c, 0.180428 \\
130 & \textcolor{LightGray}{9} & 97 & gathers, shirley, 0.180414 \\
125 & \textcolor{LightGray}{10} & 98 & good-cohn, meredith, 0.1804 \\
126 & \textcolor{LightGray}{11} & 101 & mcguffee, tyler ann, 0.180385 \\
127 & \textcolor{LightGray}{12} & 100 & edwards, sara l, 0.180369 \\
128 & \textcolor{LightGray}{13} & 102 & rom, colin, 0.180353 \\
131 & \textcolor{LightGray}{14} & 73 & deatrick, elizabeth, 0.165221 \\
152 & \textcolor{LightGray}{15} & 168 & blackburn, amy, 0.143186 \\
\bottomrule
\end{tabular}
%
% -- QDSFM
%
\begin{tabular}{*{3}{@{}p{15pt}@{}}p{106pt}@{}}
\toprule
\multicolumn{4}{c}{Seeded PageRank-HyperGraph} \\
\midrule
1 & 1 & \textcolor{LightGray}{1} & conrad, patricia, 0.202415 \\
3 & 19 & \textcolor{LightGray}{2} & folkers, greg, 0.062274 \\
2 & 26 & \textcolor{LightGray}{3} & billet, courtney, 0.051916 \\
4 & 145 & \textcolor{LightGray}{4} & collins, francis, 0.05098 \\
16 & 21 & \textcolor{LightGray}{5} & niaid odam, 0.039897 \\
6 & 5 & \textcolor{LightGray}{6} & goldner, shannah, 0.038423 \\
18 & 70 & \textcolor{LightGray}{7} & auchincloss, hugh, 0.026288 \\
5 & 85 & \textcolor{LightGray}{8} & lane, cliff, 0.023779 \\
10 & 35 & \textcolor{LightGray}{9} & routh, jennifer, 0.022279 \\
8 & 38 & \textcolor{LightGray}{10} & marston, hilary, 0.020105 \\
21 & 143 & \textcolor{LightGray}{11} & tabak, lawrence, 0.018527 \\
29 & 6 & \textcolor{LightGray}{12} & figliola, mike, 0.015333 \\
20 & 81 & \textcolor{LightGray}{13} & erbelding, emily, 0.014814 \\
15 & 40 & \textcolor{LightGray}{14} & awwad, david, 0.012929 \\
28 & 80 & \textcolor{LightGray}{15} & niaid ocgr leg, 0.011917 \\
\bottomrule
\end{tabular}
%\end{fullwidth}
    \caption{Seeded PageRank and Sparse PageRank results on a graph (left) and hypergraph (middle and right) show surprising differences among the highly ranked nodes of the diffusion -- indicating this is a useful dataset for further study. We do this for a sparse PageRank diffusion on a graph projection of the hypergraph, a direct sparse PageRank diffusion on the hypergraph, and a unregularized PageRank diffusion on the hypergraph, all seeded on Patricia Conrad.}
    \label{tab:pagerank-hypergraph}
\end{fullwidthtable}

\paragraph{Hypergraph cuts compared with graph cuts}
Hypergraph cuts can be far more interesting than simple graph cuts~\cite{Veldt-2020-hypergraph}. Here, we show how hypergraph cuts in these data are more stable. We consider the same hypergraph, but where large hyperedges are removed via a max hyperedge size filter. We see a large difference in the graph cut in the clique projected hypergraph, but relatively little difference in the hypergraph st cut between Francis Collins and Patricia Conrad (\figref{fig:hypergraph-stcut}). This is true even for multiple ways of weighting a hyperedge in the clique projection.

\begin{tuftefigure}[t]
\footnotesize
\begin{tabularx}{\linewidth}{XXX}
Maximum hyperedge size 10 &
Maximum hyperedge size 15 & 
Maximum hyperedge size no limit
\end{tabularx}

\rotatebox[origin=l]{90}{Standard clique weights}
\includegraphics[width=0.31\linewidth]{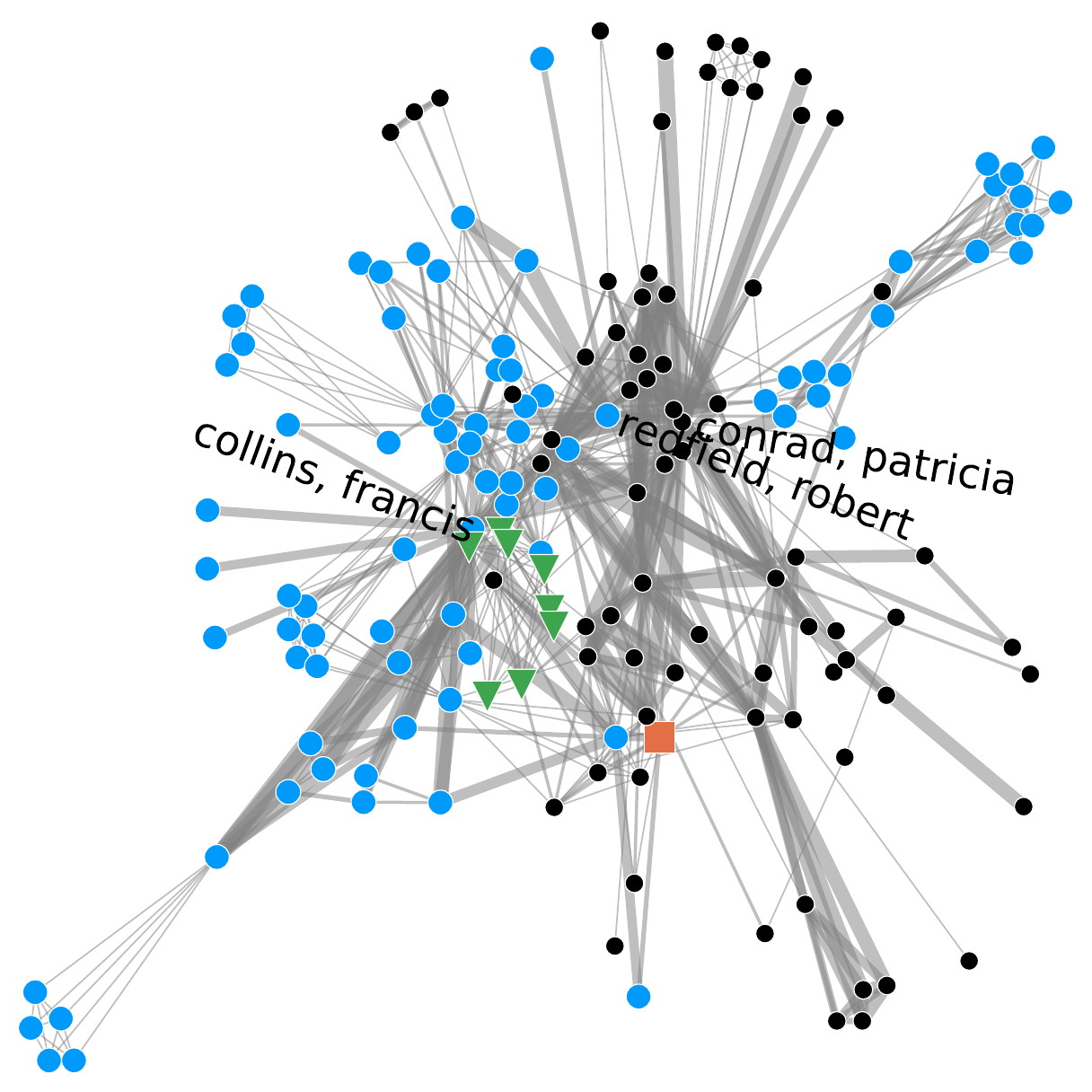}
\includegraphics[width=0.31\linewidth]{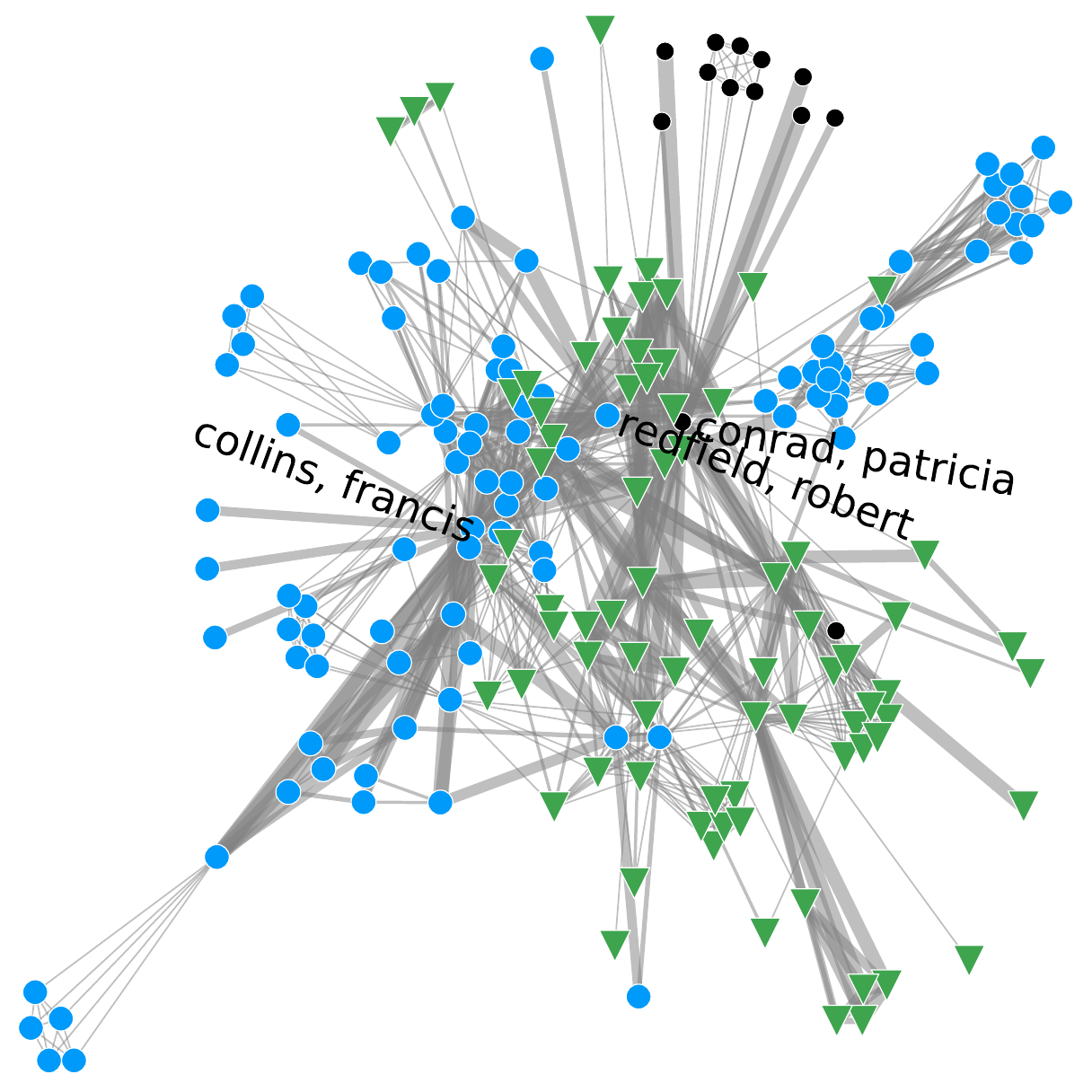}
\includegraphics[width=0.31\linewidth]{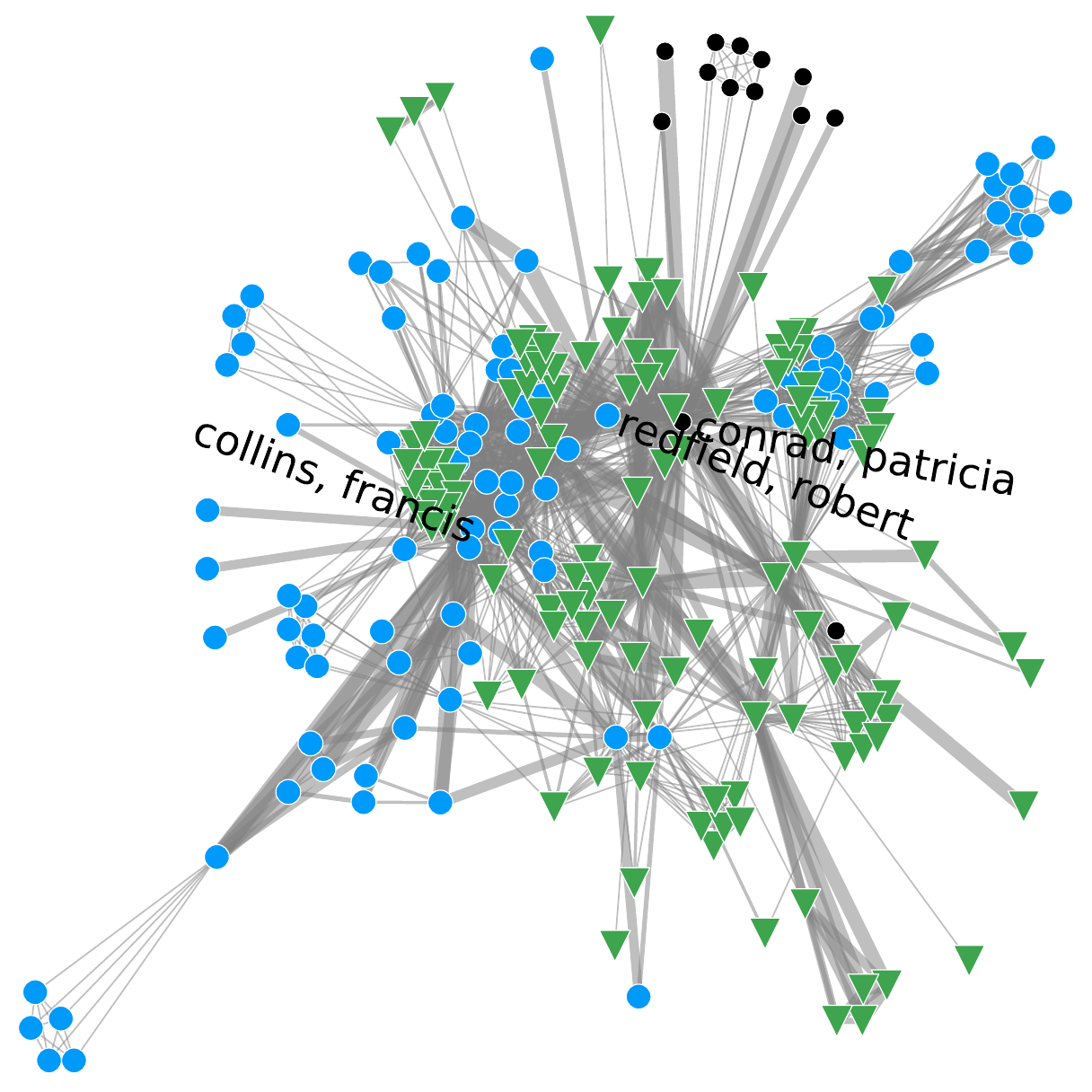}

\rotatebox[origin=l]{90}{Distributed clique weights}
\includegraphics[width=0.31\linewidth]{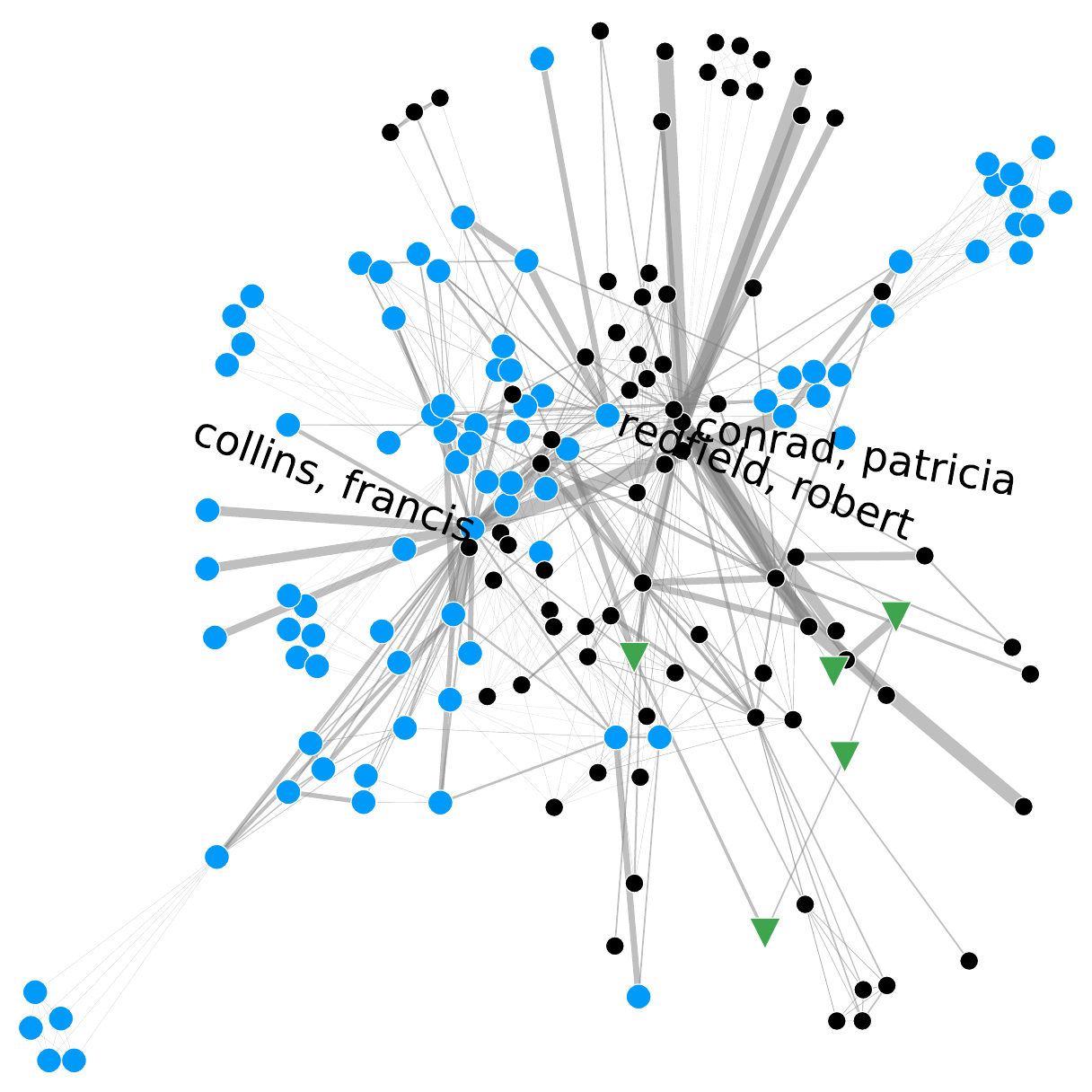}
\includegraphics[width=0.31\linewidth]{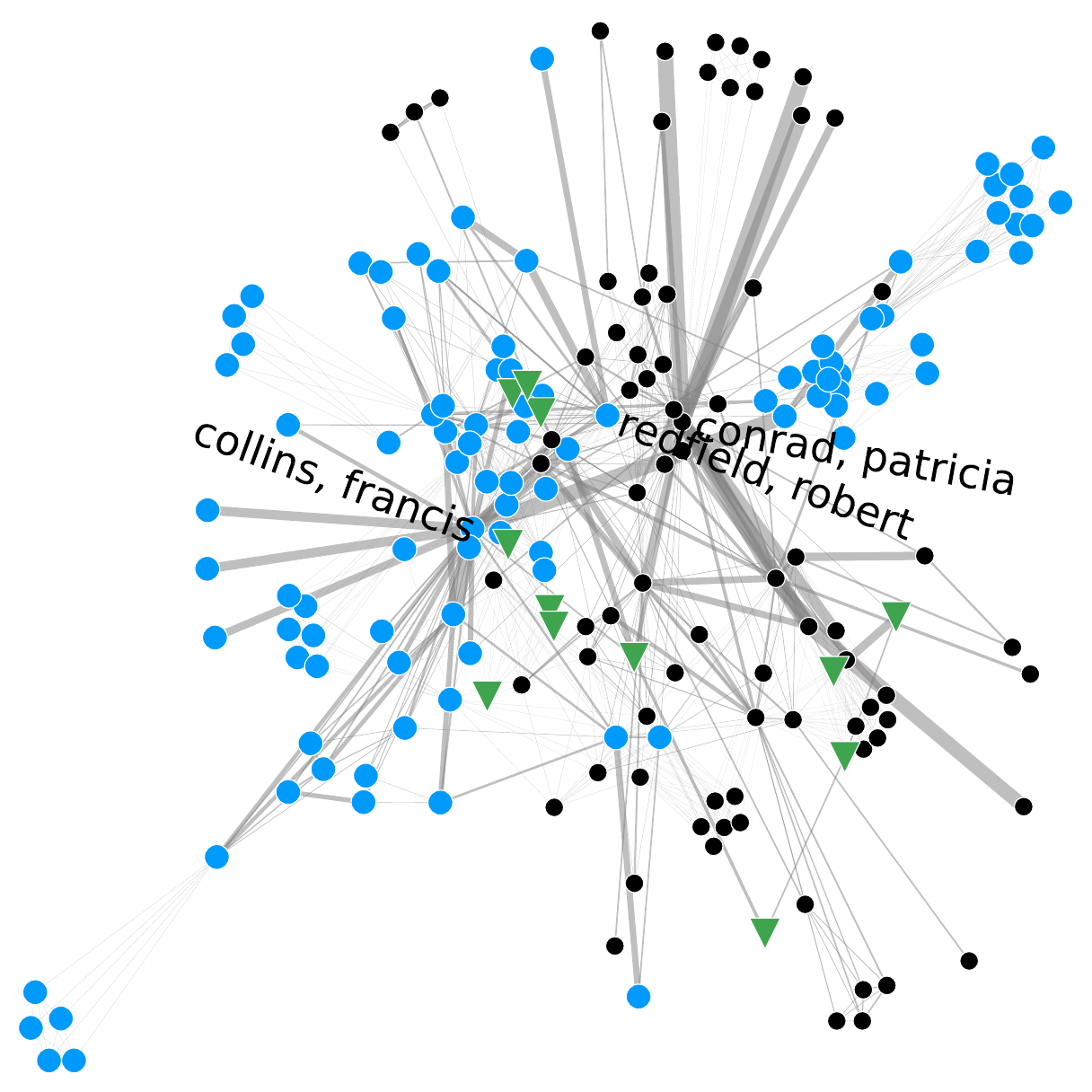}
\includegraphics[width=0.31\linewidth]{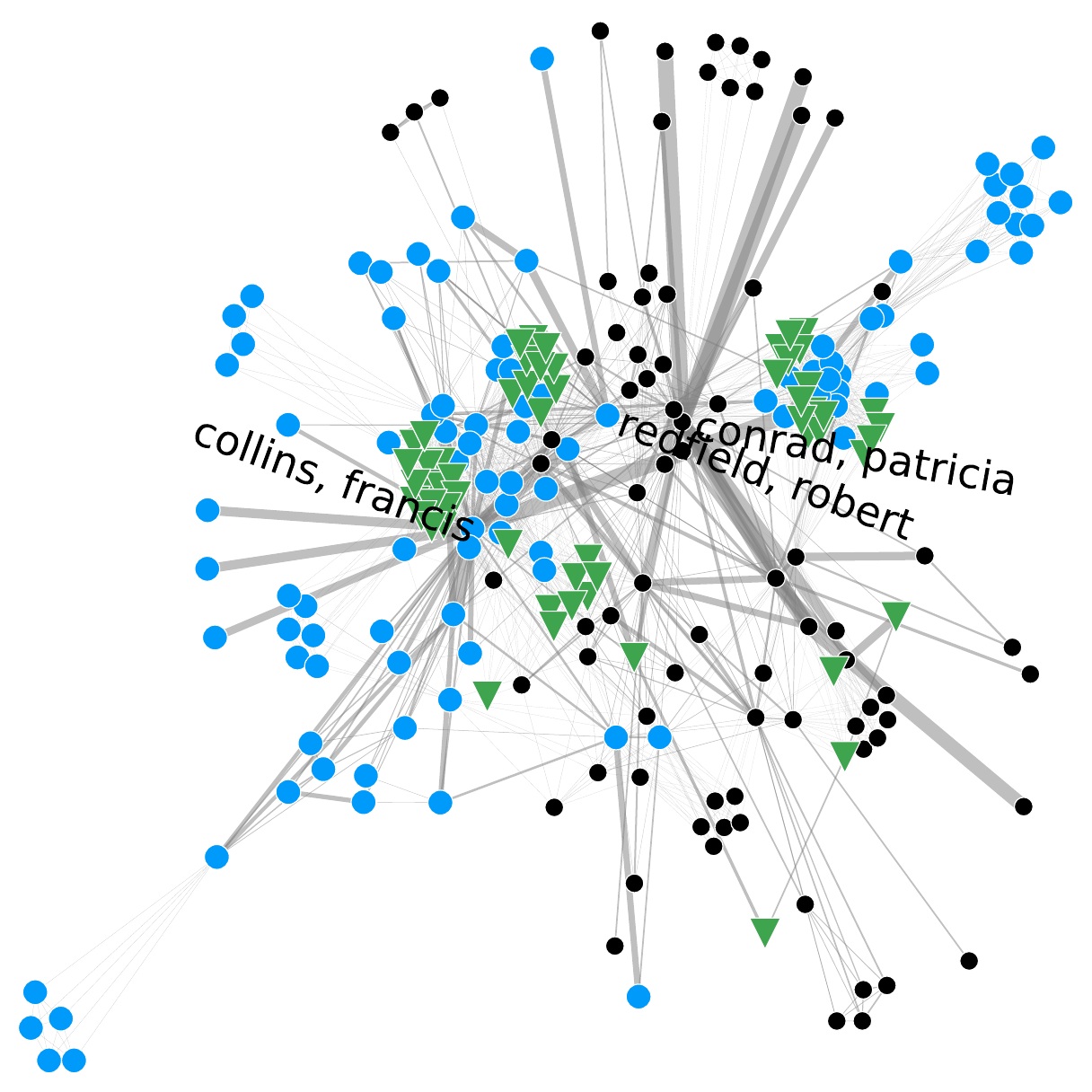}

\caption{These figures show that the hypergraph cuts are far more stable with respect to including large hyperedges compared with the graph cuts. The light blue nodes are in both the graph and hypergraph  cut between Collins and Conrad (on the Collins side). The sole light red node is in the hypergraph cut but not in the graph cut. The green nodes are in the graph cut but not hypergraph cut. (Black nodes are on the Conrad side of the cut.)  In the top row, the graph cuts are formed by projecting each hyperedge to a clique and then solving an st cut problem in the graph. If instead the graph is formed by projecting each hyperedge to a clique and weighting each edge by 1/hyperedge-size-choose-2 (so the sum of weights in the clique is 1) then we arrive at similar results with the figures in the bottom row.  Edge sizes show the various weights in the graph. Anecdotally, we note that Robert Redfield, the head of the CDC, is strongly associated with the large hyperedges that cause the graph cut to change.}

\label{fig:hypergraph-stcut} 
\end{tuftefigure}

\paragraph{Hypergraph cut flexibility}
As mentioned, hypergraph cut functions can be far more flexible than simple graph cut functions. One of the cut functions proposed by~\citet{Veldt-2020-localized-ratio} was the $\delta$-linear penalty, which interpolates between the \emph{all-or-nothing} hyperedge cut and the \emph{star-expansion} hyperedge cut function. In Figure~\ref{fig:delta}, we show nodes that switch sides while varying $\delta$ in this cut function in the hypergraph. This shows non-monotonic behavior. 

\begin{tuftefigure}[t]
\centering
\includegraphics[width=0.9\linewidth]{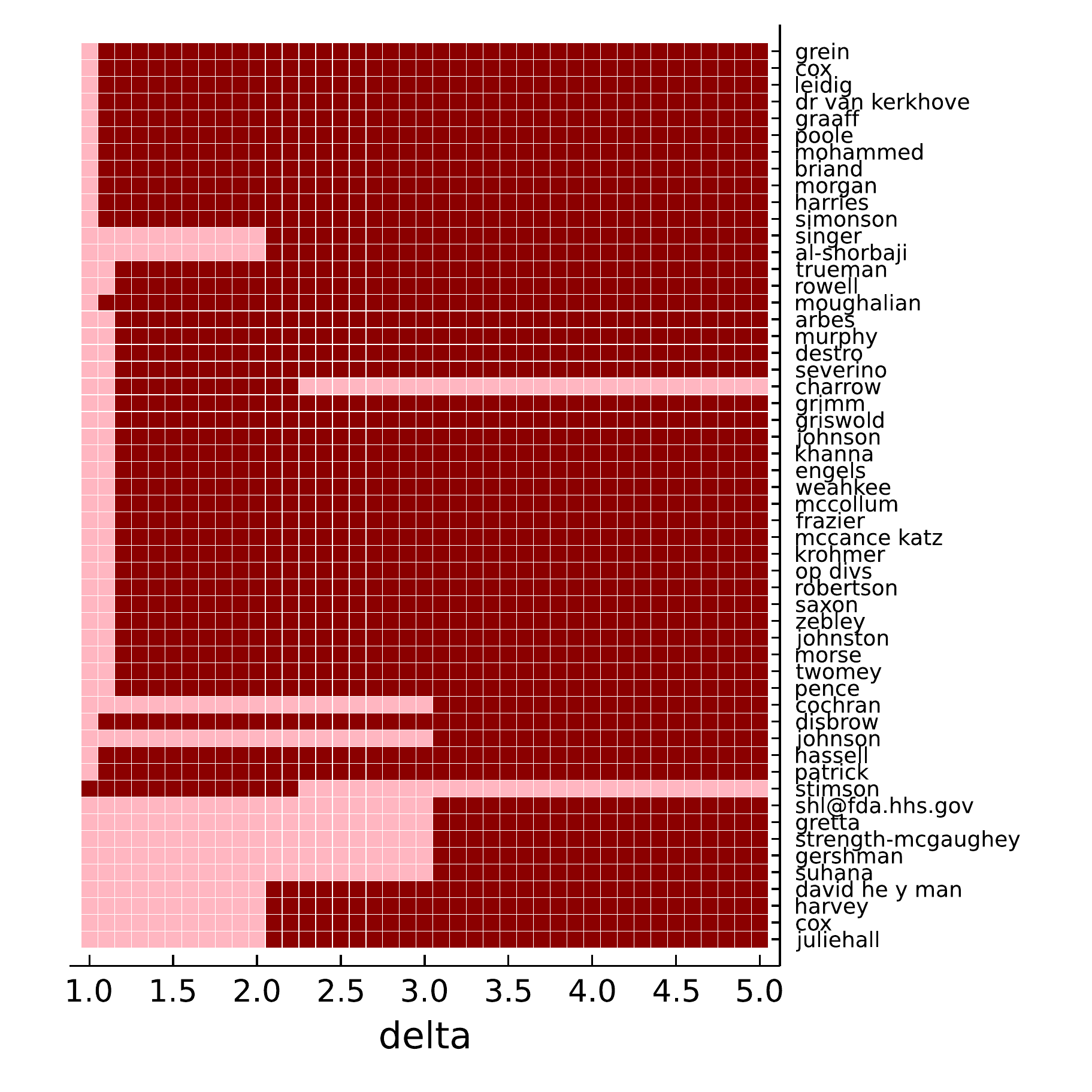}
\caption{As an example of the flexibility of hypergraph cuts, this figure shows nodes that change sides as $\delta$ is varied in a hypergraph cut between Francis Collins and Patricia Conrad. Dark red indicates the node is on the Collins side of the cut and light red is on the Conrad side. Note that the behavior is not monotonic and nodes can move back across the cut as $\delta$ increases.}
\label{fig:delta}
\end{tuftefigure}

\paragraph{Methods for local diffusions and cuts} We use the tools and codes from~\cite{Veldt-2020-hypergraph,Veldt-2020-localized-ratio,Liu-2021-localhyper} for these computations. 

%\subsection{Time series analysis}

\subsection{Temporal graph analysis}
\label{sec:temporal}
We processed the data in a set of directed edges for emails that were sent on the same day, restricted to the largest temporal strong component.\footnote{A temporal strong component~\cite{Bhadra-2003-dynamic-components,Nicosia-2012-time-varying-components} of a temporal graph is a set of nodes where there is a time-respecting path among all vertices in the component.  This gives a set of 77 nodes.} This gave a sequence of 100 adjacency matrices for each day from February 1 2020 to May 5 2020 with a few other preliminary days (e.g.~a September 4, 2018 email from Folkers to Fauci on CDC guidelines on aerosol protections for influenza and coronaviruses, Page 429). 

The first analysis we did was a temporal communicability analysis~\cite{Grindrod2011}. This analysis scores each node based on a weighted average of the length of email chains they start (broadcast-centrality) or receive (receive-centrality). The results are in \tabref{tab:temporal-communicability}.

\begin{tuftetable}[t]
\footnotesize\centering
%
% -- broadcast
%
\begin{tabular}{*{2}{p{16pt}@{}}p{112pt}}
\toprule
\multicolumn{3}{c}{broadcast} \\
\midrule
\textcolor{LightGray}{1} & 3 & fauci, anthony, 203.879418 \\
\textcolor{LightGray}{2} & 1 & conrad, patricia, 57.498332 \\
\textcolor{LightGray}{3} & 4 & billet, courtney, 49.752466 \\
\textcolor{LightGray}{4} & 29 & farrar, jeremey, 46.764788 \\
\textcolor{LightGray}{5} & 6 & collins, francis, 41.20197 \\
\textcolor{LightGray}{6} & 10 & routh, jennifer, 26.392133 \\
\textcolor{LightGray}{7} & 2 & folkers, greg, 23.055208 \\
\textcolor{LightGray}{8} & 12 & tabak, lawrence, 15.966546 \\
\textcolor{LightGray}{9} & 21 & myles, renate, 14.847189 \\
\textcolor{LightGray}{10} & 24 & lapook, jon, 13.122241 \\
\bottomrule
\end{tabular}
%
% -- receive
%
\begin{tabular}{*{2}{p{16pt}@{}}p{112pt}}
\toprule
\multicolumn{3}{c}{receive} \\
\midrule
2 & \textcolor{LightGray}{1} & conrad, patricia, 127.128824 \\
7 & \textcolor{LightGray}{2} & folkers, greg, 59.406403 \\
1 & \textcolor{LightGray}{3} & fauci, anthony, 59.291793 \\
3 & \textcolor{LightGray}{4} & billet, courtney, 44.250949 \\
20 & \textcolor{LightGray}{5} & lerner, andrea, 29.876871 \\
5 & \textcolor{LightGray}{6} & collins, francis, 29.133296 \\
19 & \textcolor{LightGray}{7} & lane, cliff, 28.915957 \\
24 & \textcolor{LightGray}{8} & cassetti, cristina, 28.679256 \\
16 & \textcolor{LightGray}{9} & marston, hilary, 27.165942 \\
6 & \textcolor{LightGray}{10} & routh, jennifer, 25.942534 \\
\bottomrule
\end{tabular}

\caption{Among the 77 nodes in the largest temporal strong component, the top 10 nodes by temporal sender and receiver centrality~\cite{Grindrod2011} with parameter $0.02$ show Fauci and Conrad as the top broadcast and receiver nodes, respectively. The light fontcolor indicates the rank in the sorted list and the dark fontcolor indicates the rank in the other list. The value after the name is the centrality score itself.   }
\label{tab:temporal-communicability}
\end{tuftetable}

The second analysis was a temporal community analysis~\cite{Mucha-2010-community}. This analysis assigns a community or group to each node at each time-point to reflect how the groups change over time. Formally, this is a modularity-like  analysis on a temporally-linked graph -- this allows the analysis to violate a strict arrow of time and foreshadow the future. The communities this analysis identifies show how the emails sent respond to various external events (\figref{fig:temporal-modularity}); although there are a few groups (i.e. the lime green around April 20th, 2020) that are harder to resolve. 

\begin{tuftefigure}[t]
    \centering
    \includegraphics[width=0.9\linewidth]{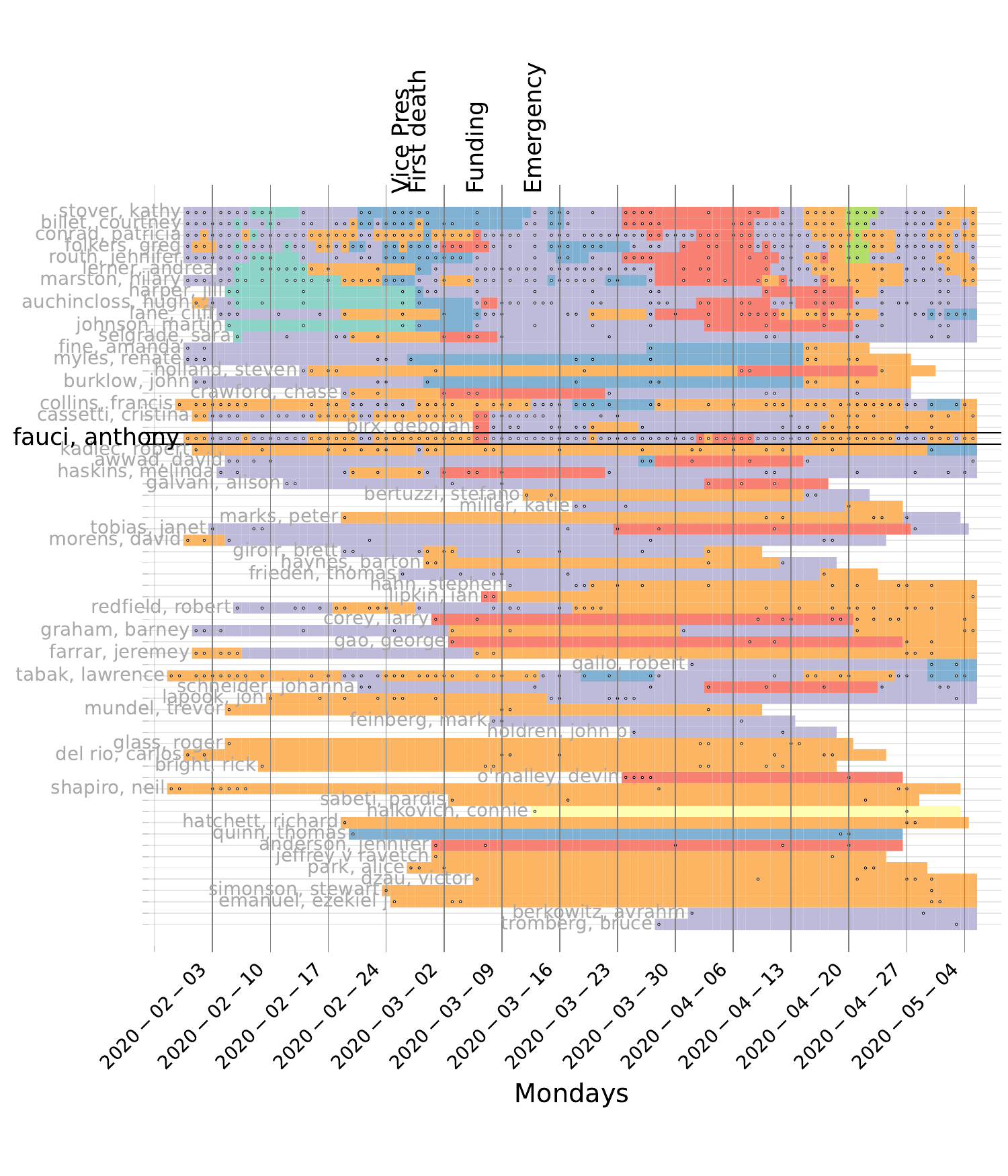}
    \caption{A plot of the communities in a temporal modularity analysis of the network; the figure should be viewed zoomed in and studied for best effect. There are 7 groups, indicated by colors.  Nodes are sorted by the number of distinct communities they are a part of, so the first few nodes switch between communities through the time-course of the emails. Community assignments are hidden until the node sends their first email and the small circles indicate days the individuals sent email along with 7 days after their last email. A few key dates are listed at top. The ``Vice Pres'' event is when Vice President Pence was appointed head of the Coronavirus Task Force; the first death of an American with COVID-19 was on Feb 28; there was a supplemental funding package passed on March 6, 2020; and there was a national emergency declaration on March 13, 2020. Fauci's node is highlighted in the middle. }
    \label{fig:temporal-modularity}
\end{tuftefigure}

We also created a force directed animation of this dataset to illustrate the temporal modularity groups. This animation is available from our github repository \url{https://github.com/nveldt/fauci-email/blob/master/figures/anim-mod.mp4}. 

\paragraph{Methods for temporal strong components} The largest temporal strong component can be computed by building a reachability network among temporal paths and then finding the largest clique in the reachability network~\cite{Bhadra-2003-dynamic-components,Nicosia-2012-time-varying-components}. We did this and used the pmc software~\cite{Rossi-2015-max-clique} to find the largest clique. This gave a set of 77 nodes.  Although the largest clique is NP-hard in general, in this case, the largest clique has the same size as the largest network core, which means it is easy to validate. Consequently, this clique can be validated by finding the largest network core and then using a greedy heuristic clique finder inside that core to find the set of 77 vertices. 

\paragraph{Methods for temporal communicability.} Let $\mA_1, \ldots, \mA_T$ be the sequence of adjacency matrices. Then the broadcast and receive temporal communicability scores are the row and column sums of the matrix $\mQ = \prod_{t=1}^T (\mI - \alpha \mA_t)^{-1}$, respectively.  The matrices involved were all small (77 nodes) and we computed this by direct inversion of the matrix -- this is in violation the pedagogical dogma of numerical linear algebra classes and would have failed the final author in Gene Golub's numerical analysis class.\sidenote[][-20ex]{The use of \texttt{inv} was because the \texttt{prod} function in Julia cannot work with a factorization object directly for successive inverses. That same author will investigate strategies in this area as this is the second time this issue has arisen in the past few months.}

\paragraph{Methods for temporal modularity} 
To compute temporal modularity, we used the Louvain algorithm directly on the \emph{slice-expanded} modularity matrix~\cite{Blondel-2008-louvain} (see reproduction details below). The modularity matrix slices were coupled with parameter 0.5, as was indicated as a reasonable default parameter in~\citet{Mucha-2010-community}.  We only briefly investigated sensitivity to this parameter and this can obviously be tuned for different effects -- we plan to explore that in the future.

\subsection{Tensor analysis}
\label{sec:tensor}

Here, we explore some higher-order structure in the emails through sender--receiver--CC interactions.
We first found a maximal set of nodes where everyone participates in the sender, receiver, and CC roles with all of the other nodes in the set.
Specifically, we examine all emails containing at least one recipient and at least one CC and find the set of discard nodes $S$ corresponding to people that do not appear at least once as a sender, receiver, and CC in these emails.
After, we discard emails where a node in $D$ is a sender, and omit nodes in $D$ from the recipient and CC lists of the other emails.
This process is repeated until there are no nodes in the discard set.
In the end, there remained a set $S$ of 44 nodes and 1,413 emails with a sender, at least one recipient, and at least one CC from $S$.

We next constructed a $44 \times 44 \times 44$ (non-symmetric) tensor $T$ representing the email relationships of the nodes $S$.
Let $s_i$ represent the sender of the $i$th email and $r_i$ and $c_i$ the subsets of $S$ who are recipients and CC.
Then the tensor entries map the total email volume the nodes, scaled by the number of email participants:
\[
T_{u,v,w}= \sum_{i} \frac{1}{\lvert c_i \rvert \cdot \lvert r_i \rvert}I(u \in c_i) I(v \in r_i) I(w = s_i),
\]
where $I(\cdot)$ is the indicator function.

Finally, we computed the hypergraph $H$-eigenvector centrality scores~\cite{benson2019three} for $T$, which is a positive unit-1-norm (unit-sum) vector $x$ such that
\[
\lambda x_u^2 = \sum_{v,w}T_{u,v,w}x_vx_w
\]
for all indices $u$ and some scalar $\lambda > 0$.
Since the first index of $T$ corresponds to CC, the centrality scores are a measure of how central each node is with respect to participation in that role
($x$ would be the same if we permuted the second and third indices, so only the first index determines the interpretation of the centrality).

\begin{margintable}
\centering\footnotesize 
\caption{Top 10 nodes in terms of CC-based tensor H-eigenvector centrality. This is 
the only list in this document where Melinda Haskins and John Mascola are top centrality nodes.}
\label{tab:hec_cc}
\begin{tabular}{r l}
\toprule
\textcolor{LightGray}{1} & conrad, patricia, 0.123106 \\
\textcolor{LightGray}{2} & folkers, greg, 0.095163 \\
\textcolor{LightGray}{3} & billet, courtney, 0.075963 \\
\textcolor{LightGray}{4} & routh, jennifer, 0.064585 \\
\textcolor{LightGray}{5} & stover, kathy, 0.061403 \\
\textcolor{LightGray}{6} & marston, hilary, 0.055949 \\
\textcolor{LightGray}{7} & haskins, melinda, 0.043622 \\
\textcolor{LightGray}{8} & tabak, lawrence, 0.043443 \\
\textcolor{LightGray}{9} & fauci, anthony, 0.037472 \\
\textcolor{LightGray}{10} & mascola, john, 0.034682 \\
\bottomrule
\end{tabular}
\end{margintable}

\Tabref{tab:hec_cc} reports the top-10 nodes in terms of this centrality measure.
Fauci is ranked ninth even though the entire dataset is constructed from his emails.
However, Fauci is in the CC position relatively less often
(Fauci was ranked first if the first index of the tensor corresponded to the sender or recipient roles).
Conrad is ranked first, which agrees with her central role in many graphs constructed from this dataset (\hyperref[tab:pagerank]{tables~\ref{tab:pagerank}} \hyperref[tab:degree]{and~\ref{tab:degree}}).
Folkers, Fauci's Chief of Staff, is ranked second.

\paragraph{Tensor text analysis}
We also release a tensor (\texttt{fauci-email-tensor-words.json}) that mirrors many analyses of the Enron email data~\cite{Cohen-2004-enron} where we examine interactions among sender, receivers, time, and words. This gives a $77 \times 77 \times 100 \times 212$ tensor of the most common words. However, we were unable to identify any useful processing of this tensor. Standard factorization analysis would often focus on individual hyperedges as the relevant factors. We leave this as a challenge for others. 

\section{Caveats}

Please remember that this not \emph{all} of Fauci's email from the relevant timeframe. We may update this document if we have more explicit documentation on what all was included or excluded in the released dataset. 

The processing of this data was automated. While we attempt to describe the major scenarios and edge-cases above and discuss how we handled them, please be aware that the information may be inaccurate. In terms of sociological findings for which they may be appropriate, these data should be used with care to understand nuances regarding the exact data collection and ingestion. 

It is very likely that additional relevant correspondence took place over the phone and text messages that are not included in the data.

Note also that the text fields of our released data have many errors. This renders text analysis problematic and we leave text analysis to future studies.

Although this data is superficially similar to the Enron data tensor frequently analyzed, there are some critical differences. First, much of the email information was redacted. Second, we only have \emph{Fauci}'s view on the email instead of raw email inbox dumps for more executives. 

\section{Outlook with this data}
We found this data extreme interesting for its seemingly unique ability to show differences among closely related methods. We have highlighted many of those features. The data is also small and easy-to-process, even with combinatorial optimization tools that are infeasible on larger data. We hope it becomes a useful resource to others as well!

\marginnote[6ex]{\emph{References deliberately pointed to arXiv versions for ease-of-access.}}
\begin{fullwidth}
\bibcolumns=3
\bibliographystyle{plainnat}
\bibliography{refs-arxiv.bib}
\end{fullwidth}

\appendix 
\section{Supplemental details}

\subsection*{Full list of derived datasets and associated files}
See \tabref{tab:dataset-files} for the files and brief associated descriptions of derived products. 

\begin{fullwidthtable}
\begin{tabularx}{\linewidth}{rX}
\texttt{fauci-email-graph.json} & the json digest of threaded emails $\cdot$ \secref{sec:processed}\\
\midrule 
\texttt{fauci-email-repliedto.json} & the \texttt{repliedto} network $\cdot$ \secref{sec:graphs}\\
\texttt{fauci-email-tofrom-5.json} & the \texttt{tofrom-nofauci-nocc} network $\cdot$ \secref{sec:graphs}\\
\texttt{fauci-email-tofrom-cc-5.json}  & the \texttt{tofrom-nofauci} network $\cdot$ \secref{sec:graphs}\\ 
\texttt{fauci-email-hypergraph-projection.json}  & the \texttt{hypergraph-proj-nofauci-nocc} network $\cdot$ \secref{sec:graphs}\\
\texttt{fauci-email-hypergraph-projection-cc.json} & the \texttt{hypergraph-proj-nofauci} network $\cdot$ \secref{sec:graphs}\\
\midrule 
\texttt{fauci-email-hypergraph.json} & the hypergraph of senders and receivers $\cdot$ \secref{sec:hypergraph} \\
\midrule 
\texttt{fauci-email-bydate-sequence-tofrom.json} & the temporal sequence of adjacency matrices $\cdot$ \secref{sec:temporal} \\
\midrule 
\texttt{cc-recipient-sender-tensor.json} & the tensor studied in \tabref{tab:hec_cc}\\
\texttt{fauci-email-tensor-words.json} & the tensor of senders, receivers, time, and words (for the top 212 words) that we did not get any meaningful analysis from \\
\midrule 
\texttt{fauci-email-repliedto-products-simple.json} & force directed layouts, modularity, conductance, and ncut partitions for the simple graph version of the network above \\
\texttt{fauci-email-repliedto-products-weighted.json} & force directed layouts, modularity, conductance, and ncut partitions for the weighted graph version of the network above \\
\texttt{fauci-email-tofrom-5-products-simple.json} & \emph{(same}) \\
\texttt{fauci-email-tofrom-5-products-weighted.json}  & \emph{(same}) \\
\texttt{fauci-email-tofrom-cc-5-products-simple.json}  & \emph{(same}) \\
\texttt{fauci-email-tofrom-cc-5-products-weighted.json}  & \emph{(same}) \\
\texttt{fauci-email-hypergraph-projection-products-simple.json}  & \emph{(same}) \\
\texttt{fauci-email-hypergraph-projection-products-weighted.json}  & \emph{(same}) \\
\textcolor{LightBlue}{\texttt{fauci-email-hypergraph-projection-cc-products-simple.json}} & \emph{Missing (computation has not completed)}\\
\texttt{fauci-email-hypergraph-projection-cc-products-weighted.json}  & \emph{(same}) \\
\end{tabularx}
\caption{The full list of derived datasets and associated files that we produce from the raw PDF dump of Fauci's email.}
\label{tab:dataset-files}
\end{fullwidthtable}

\subsection*{Network Layouts: Force directed and Modularity}
We show a few network layouts. These were computed by using the Fruchterman-Reingold layout algorithm~\cite{Fruchterman-1991-force} as implemented in igraph~\cite{Csardi-2006-igraph}. In this paper, a force directed layout is the result of applying this algorithm to the unweighted, undirected graph. We also compute modularity-biased layouts by first computing an optimal modularity partition and then densifying edges within each optimal modularity cluster. This is done in an adhoc fashion by adding uniform random edge noise within a cluster to increase the within-edge density based on the modularity partition. (This makes the graph look more like a stochastic block model that encodes the modularity partition). Then we proceed with the same layout algorithm on the edge-augmented graph. This causes the layout to show these groups more strongly than in a straightforward spring layout, although it has the potential to mislead and make groups appear more strongly than they should given the edges alone. This is why we often show both layouts. 

\subsection*{Reproducibility notes}
The github repository contains all of the scripts we used for these figures in the \texttt{final} subdirectory. For instance, the PageRank results are produced by running \texttt{pagerank-scores.jl}. We omit an index as we hope those interesting readers can easily identify the mapping between the script names and the examples in this document. As a small exception, we note the the tensor centrality analysis (\secref{sec:tensor}) is in a Python notebook \texttt{demo-cc-recipient-sender-tensor-centrality.ipynb}. 

We also feature the same examples as \text{demo} files that would be more appropriate for explanatory use as as a basis for future studies. 

The only heuristic computations which may be difficult to reproduce are the network layouts, which we sought to make as reproducible as possible by avoiding random seeds, and the Louvain-based modularity clustering~\cite{Blondel-2008-louvain}, for which we used the HyperModularity code~\cite{Chodrow-2021-hypermodularity} without the randomization techniques. Towards those ends, we provide the clustering we found as the \texttt{final/temporal-modularity-clusters.json} file.

\end{document}